\documentclass{aa}

\usepackage{graphicx}
\usepackage{txfonts}

\defcitealias{campbell13}{C13}
\defcitealias{lapenna16}{L16}
\begin{document}

   \title{NGC 6752 AGB Stars Revisited}

   \subtitle{I. Improved AGB temperatures remove apparent overionisation of \ion{Fe}{I}}

   \author{
          S.~W.~Campbell\inst{1,2,3}
          \and
          B.~T.~MacLean\inst{2,3}
          \and 
          V.~D'Orazi\inst{4}
          \and
          L.~Casagrande\inst{5}
          \and
          G.~M.~de Silva\inst{6,7}
          \and
          D.~Yong\inst{5}
          \and
          P.~L.~Cottrell\inst{8}
          \and
          J.~C.~Lattanzio\inst{2,3}       
          }

   \institute{
              Max-Planck-Institut f{\"u}r Astrophysik (MPA),
              Karl-Schwarzschild-Strasse 1,
              D-85748 Garching, Germany\\
              \email{scampbell@mpa-garching.mpg.de}
         \and
             School of Physics and Astronomy, Monash University, Clayton
             3800, Victoria, Australia
         \and 
             Monash Centre for Astrophysics (MoCA), Monash University,
             Clayton 3800, Victoria, Australia
         \and
             INAF Osservatorio Astronomico di Padova, vicolo
             dell'Osservatorio 5, I-35122, Padova, Italy   
         \and
             Research School of Astronomy and Astrophysics, Australian
             National University, Canberra, ACT 2611, Australia  
         \and
             Australian Astronomical Observatory, 105 Delhi Rd, North Ryde, NSW
             2113, Australia
         \and
             Sydney Institute for Astronomy, School of Physics, The
             University of Sydney, NSW 2006, Australia
         \and
             Department of Physics and Astronomy, University of Canterbury, Private Bag 4800, Christchurch 8140, New Zealand
             }

   \date{Received XXX; accepted XXX}

  \abstract 
{ A recent study reported a strong
  apparent depression of \ion{Fe}{i}, relative to \ion{Fe}{ii}, in the AGB
  stars of NGC 6752. This depression is much greater than that expected
  from the neglect of
   non-local thermodynamic equilibrium effects, in particular the
   dominant effect of overionisation. The
  iron abundances derived from \ion{Fe}{I} were then used to scale all
  other neutral species in the study.
}
{ 
 Here we attempt to reproduce the apparent Fe discrepancy, and investigate
differences in reported sodium abundances.
}
{ 
We compare in detail the methods and results of the recent study with those
of an earlier study of NGC 6752 AGB stars.  Iron and sodium abundances are derived using
\ion{Fe}{I}, \ion{Fe}{II}, and \ion{Na}{I} lines. We explore various
uncertainties to test the robustness of our abundance determinations.
}
{ 
We reproduce the
large \ion{Fe}{I} depression found by the recent study, using
different observational data and computational tools. Further investigation
shows that the degree of the apparent \ion{Fe}{I} depression is strongly
dependent on the adopted stellar effective temperature. To minimise
uncertainties in \ion{Fe}{I} we derive temperatures for each star
individually using the infrared flux method (IRFM). We find that the $T_{\rm{eff}}$ scales
used  by both the previous studies are cooler, by up to 100~K; such underestimated
temperatures amplify the apparent \ion{Fe}{I} depression. Our IRFM
temperatures result in negligible apparent depression, consistent with
theory. We also re-derived sodium abundances and, remarkably, found them to be unaffected
by the new temperature scale. [Na/H] in the AGB stars is consistent between
all studies. Since Fe is constant, it follows that [Na/Fe] is also 
consistent between studies, apart from any systematic offsets in Fe.
}
{ 
We recommend the use of $(V-K)$
relations for AGB stars, based on
comparisons with our individually-derived IRFM temperatures, and their inherently low uncertainties. We plan to investigate the
effect of the improved temperature scale on other elements, and re-evaluate
the subpopulation distributions on the AGB, in the next
paper of this series.
}


   \maketitle
%
%-------------------------------------------------------------------

\section{Introduction}

Due to their relatively homogeneous stellar populations, Galactic globular
clusters (GCs) have long been used to constrain stellar evolution models of
low-mass stars
(eg. \citealt{castellani68,schwarzschild70,iben71,zinn74,norris74,sweigart97,baraffe97,salaris16}). The
colour-magnitude diagrams of GCs generally exhibit well-populated sequences
corresponding to most phases of stellar evolution. Additionally, most GCs
are chemically homogeneous in heavy elements, for example the star-to-star
variation of iron is usually smaller than the observational
uncertainties. Main sequence observations indicate that the age differences
between the subpopulations are undetectable, or small ($\lesssim 10^{8}$
years), as compared to total ages of up to 13 Gyr
(eg. \citealt{piotto07}). For most purposes, the stars in each
GC can be considered coeval.

In contrast to this homogeneity, observations of the light elements have
revealed a consistent picture of subpopulations within each GC. Supported
also by photometry, these subpopulations are most clearly seen in
multi-dimensional chemical space, for example in the Na-O plane. Indeed
\cite{carretta10} suggest that a negative/anti-correlation in Na-O is the
defining feature of a GC, clearly separating them from open clusters which
show light element homogeneity (\citealt{desilva09,maclean15}). In addition to the
well studied variation in the elements such as C, N, O, Na, Mg, Al, there
is a growing body of evidence that helium also varies
(eg. \citealt{dantona05,milone15,valcarce16}). This is qualitatively
consistent with proton-capture nucleosynthesis, whereby H is burned to He
through the CNO cycle (converting C and O to N), and Al and Na are produced
through the Mg-Al and Ne-Na cycles. We refer the reader to \cite{gratton12}
for a complete review of multiple populations in globular clusters.

Whilst the now well-established existence of multiple subpopulations in GCs
adds significant complexity to understanding GCs and their formation, it
opens up new opportunities in constraining stellar evolution models since each GC
has (at least) two populations practically identical in age and heavy
element composition, but different in light element composition. Thus GCs
can provide differential comparisons between models of different initial
light element constitutions, and in particular, helium content, which is a
dominant factor in a star's evolution
(eg. \citealt{dantona02,karakas14,chantereau16}).

Until recently chemical abundance studies of the GC multiple populations
have mainly focused on red giant branch (RGB) stars. Studies of earlier
phases of evolution such as the main sequence and sub giant branch have
shown that the proportions of stars making up each subpopulation within a
GC are generally constant through the colour-magnitude diagram. It has also
been shown that the subpopulations occupy different locations on the
horizontal branch (HB; see eg. \citealt{marino11,gratton15}).

The phase of evolution directly after the HB, the asymptotic giant branch
(AGB), has only recently started to be investigated systematically. The AGB
is particularly interesting because it should contain information about the
previous phase, the HB, which is one stage of evolution that is predicted
to diverge significantly between He-rich and He-normal stars\footnote{Due
  to the more rapid evolution of He-rich stars they have lower stellar
  masses at a given age. Since the HB core masses do not change
  significantly between He populations, the envelope masses on the HB are reduced, and
  the $T_{\rm{eff}}$ increased, giving rise to blue extensions in the observed
  HBs.}. Evolutionary models of HB stars are also known to have very
substantial uncertainties
(eg. \citealt{constantino15,campbell16,constantino16}).  Early
low-resolution spectroscopic work on GC giant stars sometimes contained a
few AGB stars (usually tentatively identified, see \citealt{campbell06conf}
for a summary). In some cases these early studies showed possible
differences in subpopulation ratios between the AGB and RGB. For example
\cite{norris81} found a lack of AGB stars with strong cyanogen (CN) band
strengths in NGC 6752, as compared to their RGB sample, and \cite{mallia78}
found a dominance of CN-strong stars on the AGB of 47 Tuc. Cyanogen
(roughly) tracks N content, such that CN-weak stars are first
generation/subpopulation (hereafter SP1) and CN-strong stars are second
subpopulation (hereafter SP2). These studies were however hampered by low
resolution, imprecise photometry (required for separating the RGB and AGB
stars), and small samples of AGB stars. Decades later the quality of
photometry had improved such that \cite{sneden00conf} and
\cite{campbell06conf} argued that it should now be possible to study the
AGB stars of GCs in a systematic way. \cite{campbell10conf} presented some
early CN results for a systematic study of AGB stars in 9 GCs, based on
medium resolution spectra ($R \sim 3000$). The findings were mixed, with a
range of interpretations being possible. This was due to the uncertainties
in molecular band formation, which is dependent on temperature, as well as
the interrelated abundances of C, N, O. One GC did appear to be a clear
case though -- NGC 6752. Its AGB was dominated by CN-poor giants, in
agreement with \cite{norris81}.  \cite{norris81} had speculated that this
may imply that all of the SP2 stars avoid the AGB phase. This is however
not expected from stellar theory -- about 50\% of the AGB stars are
predicted to be SP2 (CN-strong, Na-rich; \citealt{cassisi14}). Such a claim
of strong discordance between observation and theory required stronger
evidence. This was provided by \cite{campbell13} with sodium abundance
measurements from high-resolution spectroscopy of 24 RGB and 20 AGB stars in
NGC 6752.  The Na results confirmed the CN results, and \cite{campbell13}
inferred that all of the Na-rich (SP2) stars were avoiding the AGB phase in
NGC 6752.

Since the NGC 6752 study a number of research groups have investigated the
AGB stars of many other GCs, with high-resolution spectroscopy --
\object{47 Tuc}: \cite{johnson15}; M2, M3, M5, M13:
\cite{garciahernandez15}; M62: \cite{lapenna15}; M4: \cite{maclean16}; NGC
2808: \cite{wang16}; NGC 6752: \cite{lapenna16}. So far no consistent
picture of subpopulation ratios on the AGB has emerged. Interestingly, for
the two GCs that have been studied more than once so far, conflicting
evidence has been reported. In the case of \object{M4} photometric
inferences of population proportions (\citealt{lardo17}) disagree with the
spectroscopic results (\citealt{maclean16}). The conflicting spectroscopic
evidence for the other case, NGC 6752 (\citealt{campbell13,lapenna16}) is
the topic of the current study. Adding to the debate, a photometric
  study on \object{NGC 6752} AGB stars has very recently been accepted for
  publication (\citealt{gruyters17}). We refer the reader to
\cite{maclean16} for a more detailed summary of the literature thus far.

%--------------------------------------------------------------------
\subsection*{Conflicting results for NGC 6752}

The \cite{campbell13} study (hereafter \citetalias{campbell13}) found that
the sodium abundances in their sample of NGC 6752 AGB stars were consistent
with a single value -- the standard deviation of [Na/Fe] was $\sigma =
0.10$, comparable to the internal errors of $\sim\pm 0.1$ dex. They
reported that the single value corresponds to that of the O-rich/Na-poor
subpopulation of NGC 6752 (SP1, often referred to as `first
generation'). We note that there have been at least three subpopulations identified
in NGC 6752, one with field-star-like composition, and the other two with
enhanced Na and reduced O (amongst other light element variations, see
\citealt{carretta12}). For simplicity we divide them here into just two
groups: SP1 and SP2.

In contrast to the \citetalias{campbell13} result, \cite{lapenna16}
(hereafter \citetalias{lapenna16}) report a distinctly different
distribution in [Na/Fe]. In particular they find that about $50\%$ of their
sample have enhanced [Na/Fe] -- and corresponding (anti-) correlations with
[C, N, O, Al/Fe] (see their Fig.~2). As they state, this result is much
more consistent with current stellar evolution predictions
(\citetalias{campbell13}; \citealt{cassisi14}). \citetalias{lapenna16}
re-observed the AGB stellar sample of \citetalias{campbell13} (20 stars)
with a different instrument, the UVES spectrograph on the
VLT. \citetalias{campbell13} used data collected using FLAMES
(\citealt{pasquini03}), also on the VLT. Thus the spectra and analysis
methods are independent, but the AGB stellar samples are identical.

The \citetalias{lapenna16} study did not investigate why the results
of the two studies differ so much. The study also did not
observe or homogeneously re-analyse RGB stars, which are very useful as a
control sample, since they have been well studied in NGC 6752, and they
show the full range of Na-O dispersion for the particular analysis
methodology that one adopts (\citetalias{campbell13} included 24 RGB
stars). Here we explore the methods, uncertainties, and assumptions of both
studies with an aim to finding a robust result for [Na/Fe]. We will
investigate other elements in the next paper of the series. We begin by
directly comparing the key parameters and results of the two studies.

\section{Comparisons between \citetalias{campbell13} and \citetalias{lapenna16}}

\subsection{Stellar parameter comparison}

   \begin{figure*}
   \centering
   \includegraphics[width=0.32\linewidth]{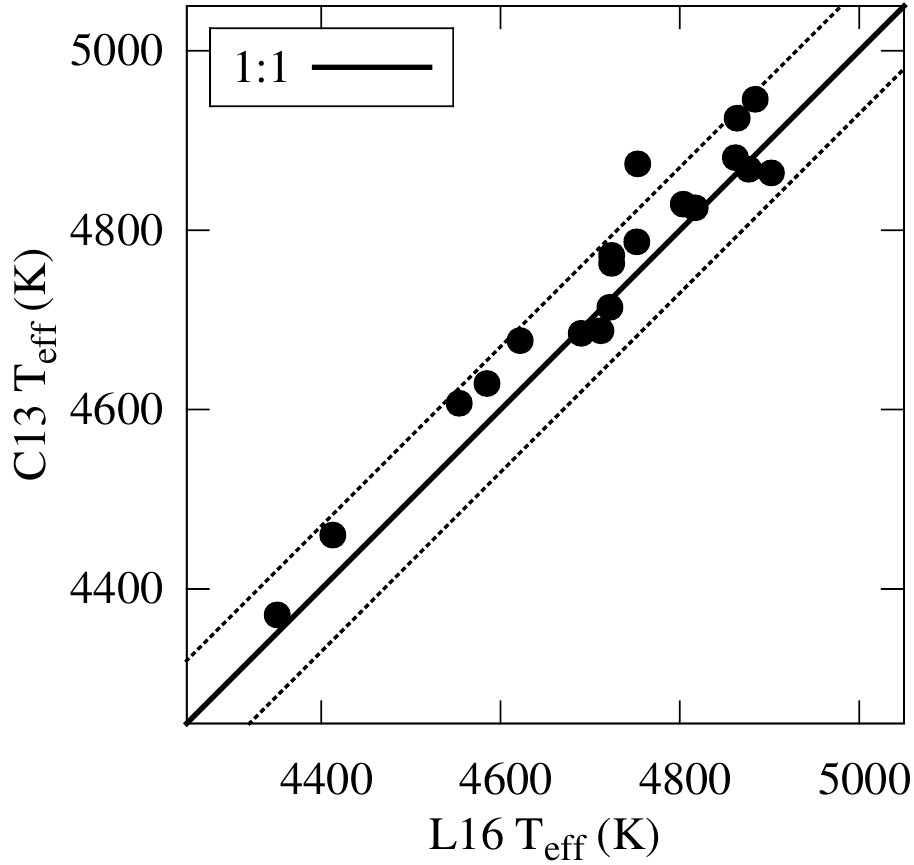}
   \includegraphics[width=0.32\linewidth]{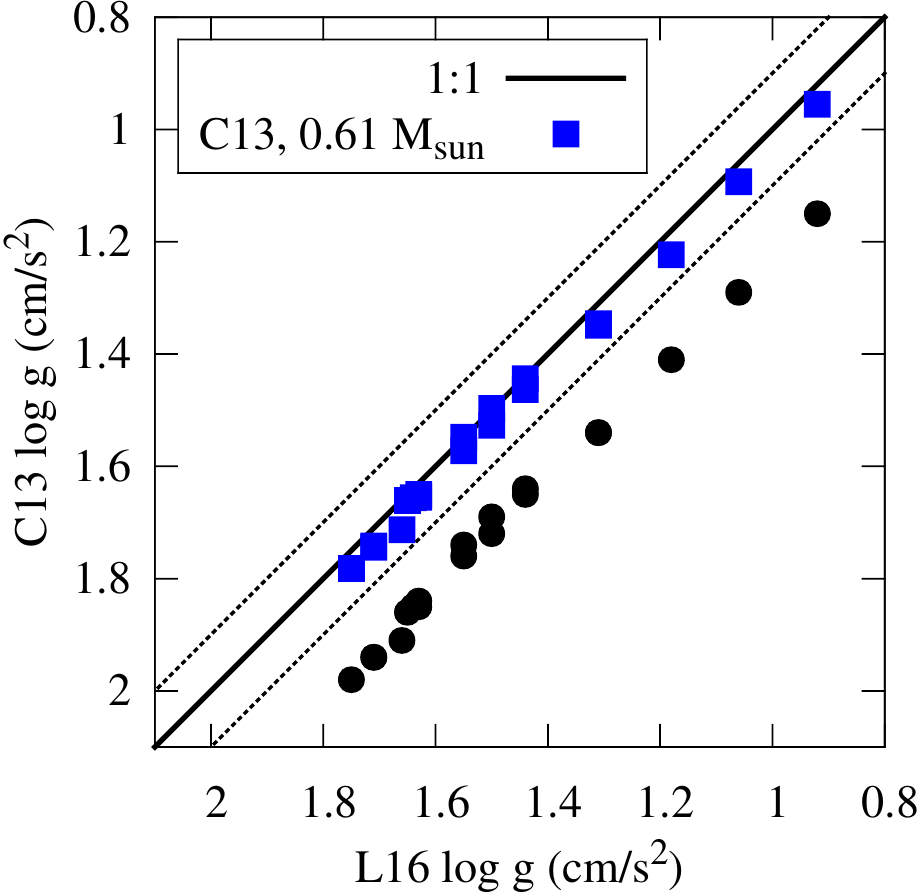}
   \includegraphics[width=0.32\linewidth]{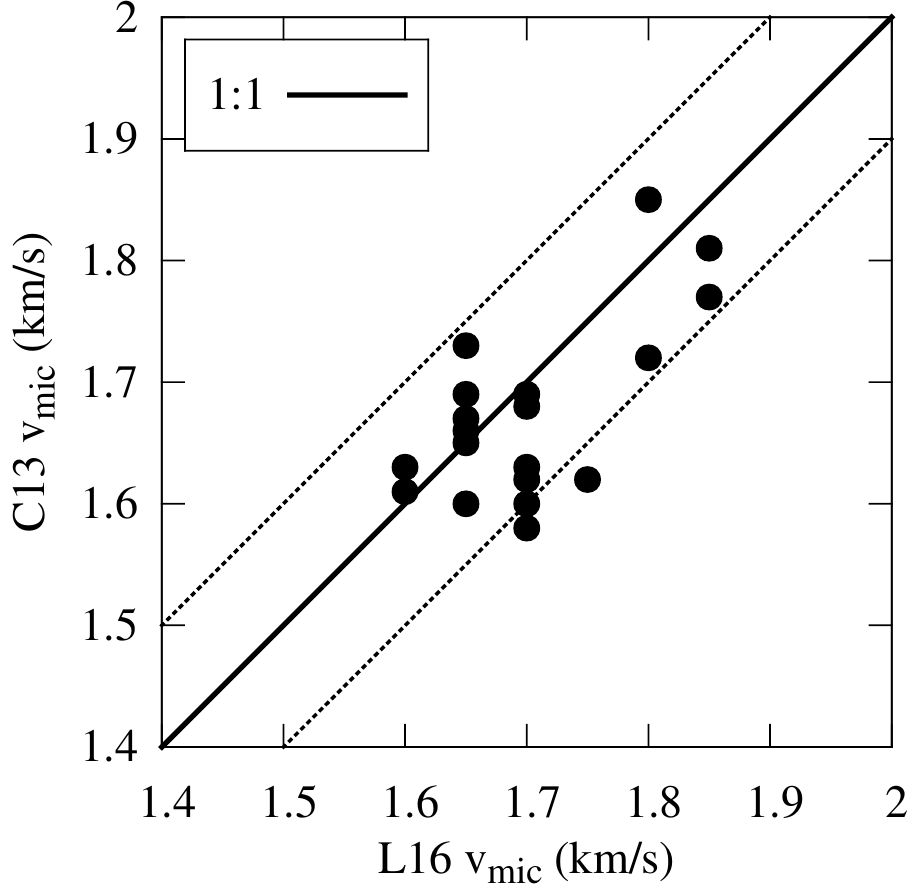}
   \caption{Comparison of $T_{\rm{eff}}$ (left), $\log g$ (centre), and
     $v_{mic}$ (right) values adopted for the AGB sample by
     \citetalias{campbell13} and \citetalias{lapenna16}. Dotted lines show
     typical uncertainties in each of the parameters ($T_{\rm{eff}}$, from
     colour-$T_{\rm{eff}}$ relation: $\pm 70$ K; $\log g$: $\pm 0.1$ km/s;
     $v_{mic}$: $\pm 0.1$ dex). The centre panel also shows our new $\log g$ values calculated
     using a more appropriate mass estimate for the AGB stars (0.61~M$_{\odot}$,
     blue squares; see text for details).}
   \label{fig:paramcompare}
   \end{figure*}

The stellar parameters -- effective temperature $T_{\rm{eff}}$, surface
gravity $\log g$, microturbulent velocity $v_{mic}$, and global metallicity
[M/H] -- are central to the spectroscopic determination of abundances. They
are the parameters that define the stellar atmosphere model one uses to
infer the strength of each line. The parameters are well known to have
degeneracies, for example a change in $\log g$ can mimic a change in
[Fe/H]. It is therefore imperative to compare the stellar parameters of
\citetalias{campbell13} and \citetalias{lapenna16}\footnote{These
  comparisons are in the context of 1D LTE abundance analyses.}.

\citetalias{lapenna16} derived $T_{\rm{eff}}$ by requiring no trend between
iron abundances and excitation potential, which is usually referred to as
`spectroscopic' $T_{\rm{eff}}$. On the other hand \citetalias{campbell13}
used `photometric' $T_{\rm{eff}}$, which is derived from
colour-$T_{\rm{eff}}$ relations. The left panel of
Figure~\ref{fig:paramcompare} shows that the $T_{\rm{eff}}$ values compare
well, with virtually all temperatures being the same within the uncertainties given
by the colour-$T_{\rm{eff}}$ relations (\citetalias{campbell13} used the
\citealt{alonso99} relations). It is interesting that there is agreement
despite the different methods used to arrive at the final temperatures
(although see Sec.~\ref{sec:ramifications}).

The centre panel of Figure~\ref{fig:paramcompare} shows that there is a
constant offset of about 0.2 dex in surface gravity ($\log g$) between
\citetalias{campbell13} and \citetalias{lapenna16}, with the
\citetalias{lapenna16} gravities being lower. This was noted by
\citetalias{lapenna16}, who suggested that it could be due to the adopted
distance modulus or stellar mass. \citetalias{campbell13} used the same
distance modulus as \citetalias{lapenna16} ($(m-M)_{V} = 13.13$;
\citealt{harris96}). However \citetalias{campbell13} neglected to account
for mass loss between the RGB and AGB. They adopted the same mass for AGB
stars as used for the RGB stars, 0.84 M$_{\odot}$, which is clearly
incorrect. Following \citetalias{lapenna16} we adjusted the mass for the
AGB stars to 0.61 M$_{\odot}$, the median HB mass inferred for NGC 6752 by
\cite{gratton10}, and recalculated $\log g$. It can be seen that this
removes the offset between \citetalias{lapenna16} and
\citetalias{campbell13}, bringing the gravities in to near perfect
agreement (blue squares in  Fig.~\ref{fig:paramcompare}).

In \citetalias{campbell13} the microturbulent velocity $v_{mic}$ was
determined using the relation of \cite{gratton96}, whilst in
\citetalias{lapenna16} it was obtained spectroscopically, by requiring no
trend between the reduced equivalent widths and abundances derived from
\ion{Fe}{i} lines. The right panel of Figure~\ref{fig:paramcompare} shows
that the \citetalias{lapenna16} values are quantised, in 0.05 km/s
steps. This is most likely due to 0.05 km/s steps being taken to arrive at
a spectroscopic solution, a reasonable approach given the uncertainty in this
parameter. The values cover a small range (1.60 to 1.85 km/s), and the two
studies agree considering the characteristic uncertainty of $\pm 0.1$ km/s.

With regards to the global metallicity used for atmospheric modelling,
\citetalias{lapenna16} used $\rm{[M/H]} = -1.50$ whilst
\citetalias{campbell13} used $\rm{[M/H]} = -1.54$. This is a small
difference and is not expected to affect the results significantly.

In summary, apart from the gravity offset, all other stellar parameters
show no significant difference between the two studies. Amongst the species
under investigation here (\ion{Fe}{i}, \ion{Fe}{ii}, \ion{Na}{i}), gravity
should mainly affect \ion{Fe}{II}. \ion{Na}{I} is expected to be largely
unaffected\footnote{This is due to \ion{Na}{I} being the minority species
  (in these stars sodium is predominantly in the form of \ion{Na}{II}) and
  thus its line formation is insensitive to pressure. Since the atmospheric
  pressure is primarily determined by gravity, it follows that the formation of
  \ion{Na}{I} lines is not sensitive to changes in gravity
  (see eg. the discussion in Chapter 13 of \citealt{gray05})\label{foot}.} and for this reason
we continue with the comparison using the published \citetalias{campbell13}
Na abundances.
\addtocounter{footnote}{-1}

\subsection{[Na/H] comparison}\label{sec:nah}

   \begin{figure}
   \centering
   \includegraphics[width=\hsize]{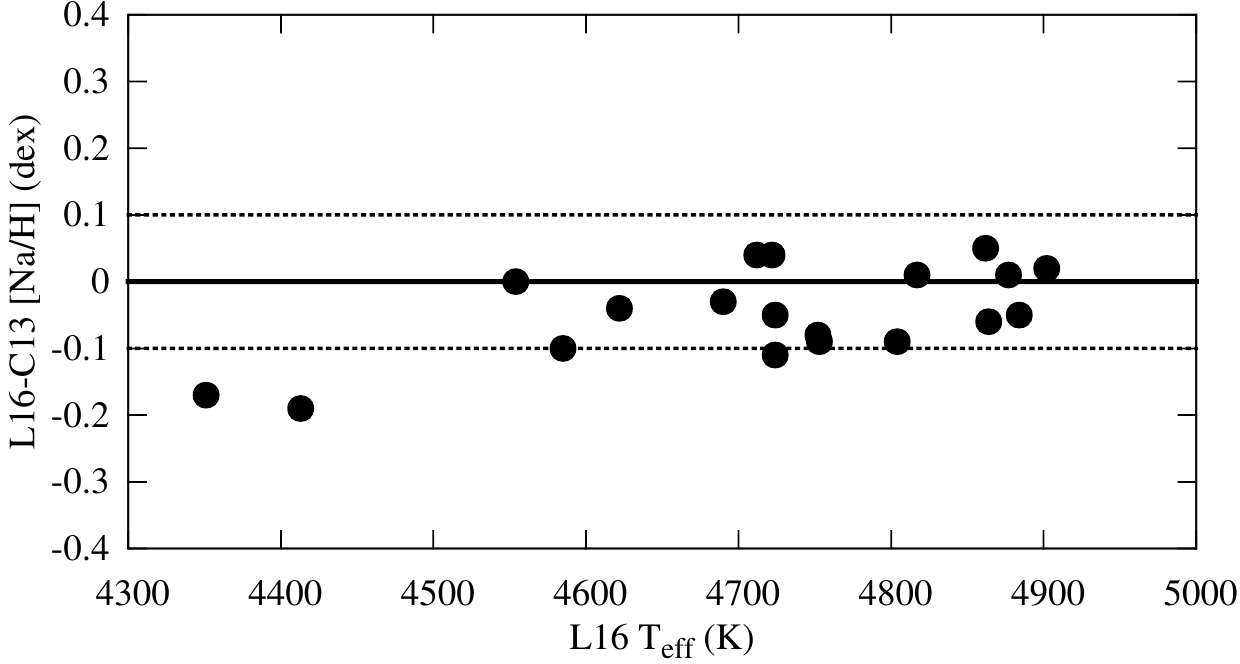}
   \caption{Difference in [Na/H] between the two studies. Dotted lines
     indicate typical uncertainties.}
   \label{fig:NaHcompare}
   \end{figure}

To derive abundances of sodium both studies used the equivalent width (EW)
method. \citetalias{campbell13} utilised the strong Na I doublets at
5682/5688 \AA~and 6154/6160 \AA, although only the first doublet was
usually measurable in the AGB stars. As far as we are aware
\citetalias{lapenna16} used the same doublet for their AGB sample (see
their Fig.~1). The \citetalias{lapenna16} data has a moderately higher
resolution (UVES, $R = 40,000$) than that obtained by
\citetalias{campbell13} (FLAMES, $R = 24,000$).

In Figure~\ref{fig:NaHcompare} we show the difference between the
\citetalias{campbell13} and \citetalias{lapenna16} [Na/H]
results\footnote{We adopt the \cite{GS98} solar Na value $\log\epsilon =
  6.33$ for scaling.}. Apart from the two coolest
stars, which have {\it lower} Na in \citetalias{lapenna16} (we discuss
these stars further in Sec.~\ref{sec:newsodium}), there is no significant
difference in [Na/H]. There is a slight systematic offset to lower [Na/H]
in \citetalias{lapenna16} ($\sim -0.05$ dex). Adding in the uncertainties
from \citetalias{lapenna16}, and considering that the uncertainties quoted
are {\it internal} only, the agreement is remarkable. This strongly
suggests that a range of factors have no significant effect on the
\ion{Na}{I} abundance derivation, including the following:
\begin{itemize}
   \item Gravity offset in \citetalias{campbell13} (as
     expected, see footnote \ref{foot})
   \item Increase in resolution in \citetalias{lapenna16} over \citetalias{campbell13}
   \item Small differences in model atmospheres and their inputs (eg. [M/H])
   \item Small differences between the spectroscopically derived
     temperatures (\citetalias{lapenna16}) and the photometric temperatures (\citetalias{campbell13})
   \item Scatter in the microturbulent velocities
\end{itemize}

The result of this comparison is reassuring and gives confidence in the
methods used to derive [Na/H].  \citetalias{campbell13} argued that their
Na results are consistent with a single value, given the uncertainties, and
that the value corresponds to SP1 of NGC~6752. This conclusion is however
at odds with the \citetalias{lapenna16} study, which concluded that the
slightly larger spread found for [Na/H] ($\sigma = 0.13$ dex versus 0.10
dex in \citetalias{campbell13}) is significant. Based on the uncertainty
estimates of \citetalias{lapenna16} ($\sim 0.06$ dex, judging from [Na/H]
in their Fig.~2), which are somewhat smaller that those of
\citetalias{campbell13} ($\sim 0.10$ dex), this conclusion may be correct
-- assuming the \citetalias{lapenna16} error estimates are realistic. We
explore various sources of uncertainty in Section~\ref{sec:realornot} and
Section~\ref{sec:uncerts}. We now investigate the considerable differences
in [Na/Fe] between the two studies.

\subsection{[Na/Fe] comparison}\label{sec:NaFeCompare}

   \begin{figure}
   \centering
   \includegraphics[width=\hsize]{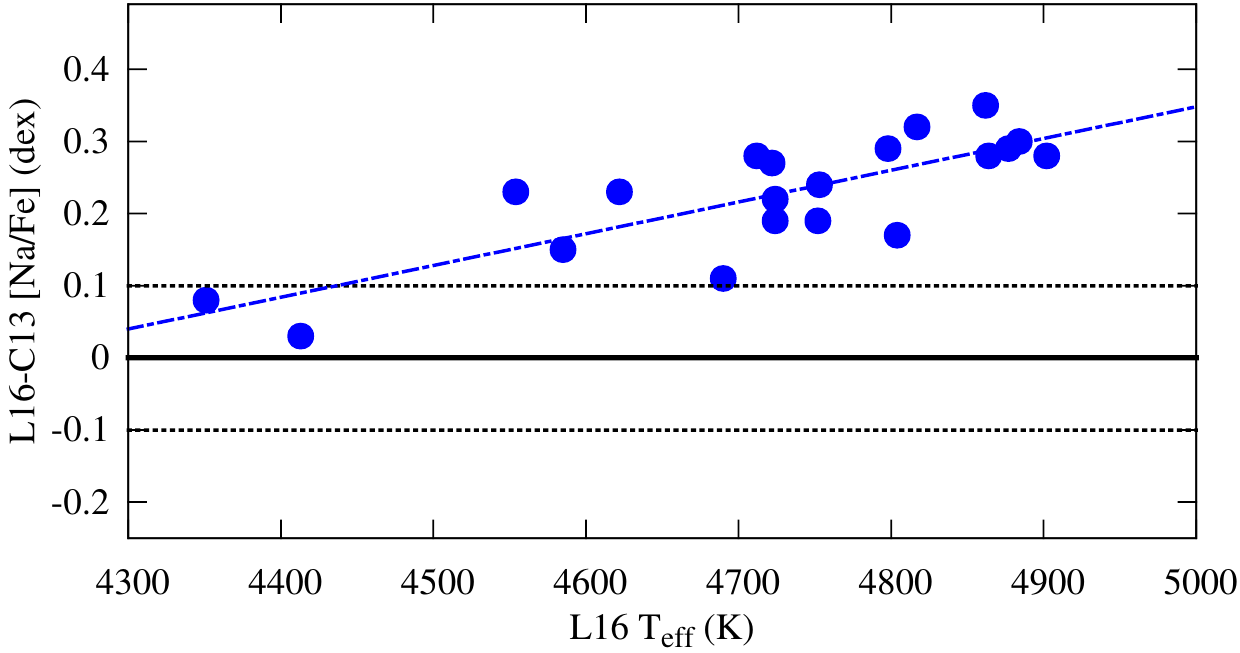}
   \caption{Difference in [Na/Fe] results between \citetalias{lapenna16}
     and \citetalias{campbell13}. The dash-dotted line is a linear
     fit. Differences much larger than the typical uncertainties (dashed
     lines) are present.}
   \label{fig:NaFecompare}
   \end{figure}

In Figure~\ref{fig:NaFecompare} we show the difference in [Na/Fe] between
the two studies. Differences of up to $+0.35$ dex can be seen, although
they range from zero to this very high value. Interestingly there is a
temperature trend, with stars with the highest $T_{\rm{eff}}$ having the
largest differences in [Na/Fe]. A linear regression analysis shows that the
Pearson correlation coefficient $r^{2} = 0.63$ and that the slope is
highly significant ($t$-statistic = $5.6 \sigma$). This was described in
\citetalias{lapenna16} as a systematic offset of 0.25~dex.

   \begin{figure}
   \centering
   \includegraphics[width=\hsize]{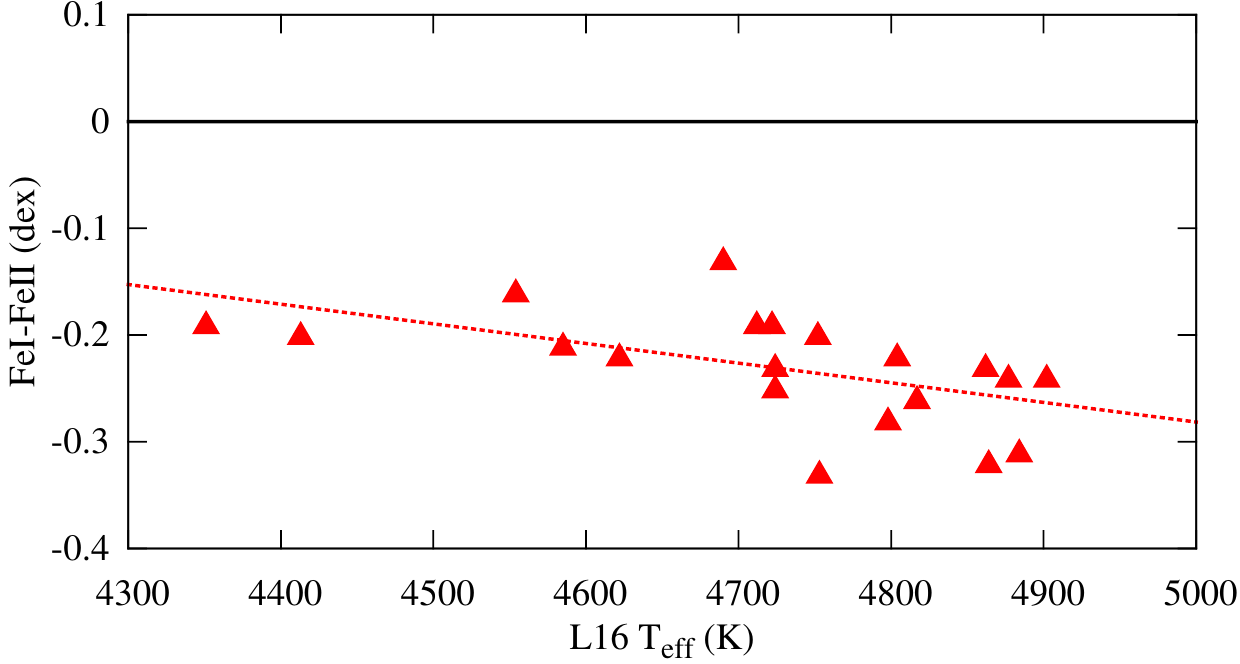}
   \caption{Depression of \ion{Fe}{I} relative to
     \ion{Fe}{II} in the \citetalias{lapenna16} study. The dashed line is a
   linear fit.}
   \label{fig:FeIdepression}
   \end{figure}

Given our conclusion about [Na/H], that the results are practically
identical between studies, the [Na/Fe] differences must be wholly driven by
the denominator, i.e. the Fe distribution must give rise to the difference
in [Na/Fe] distribution.

With respect to the methodology used to derive [Na/Fe] for the NGC 6752 AGB
stars the two studies diverge considerably. \citetalias{campbell13} did not
derive Fe abundances. They instead assumed a single Fe abundance for all
stars in their sample ($\rm{[Fe/H]} = -1.54$, \citealt{carretta07}). This
assumption is discussed further at the beginning of
Section~\ref{sec:FeC13}.  In contrast, \citetalias{lapenna16} did derive Fe
abundances, based on both \ion{Fe}{I} and \ion{Fe}{II}. An important part
of their methodology was that they did not derive $\log g$
spectroscopically, at least not in the common meaning of spectroscopically
(we set out the steps in their method in Sec.~\ref{sec:FeC13}). This was
done specifically to avoid `forcing' the abundances of \ion{Fe}{I} and
\ion{Fe}{II} to be equal. To motivate this choice \citetalias{lapenna16}
cite some studies that have reported Fe differences, $\delta\rm{Fe} =
\ion{Fe}{I} - \ion{Fe}{II}$, in globular cluster RGB and AGB stars
(\citealt{ivans01,lapenna14,lapenna15,mucciarelli15}). Certainly not
requiring that \ion{Fe}{I} = \ion{Fe}{II} is necessary for detecting any
possible $\delta\rm{Fe}$, which would most likely be due to overionisation
of \ion{Fe}{i} (\citealt{lind12}), but, as we show later
(Sec.~\ref{sec:realornot}), care is required in order to be confident of
the quantitative results. In particular there needs to be a high level of
confidence in the stellar parameters used, otherwise an {\it apparent
  overionisation} can be misinterpreted as a real physical
phenomenon\footnote{That is, a physical phenomenon that is not captured by
  the LTE treatment.}.

Crucially, to obtain [Na/Fe] \citetalias{lapenna16} decided to use only
Fe abundances derived from \ion{Fe}{I} lines in the denominator,
following the original suggestion of \cite{ivans01} (see also
\citealt{lapenna14,lapenna15,mucciarelli15}). Moreover, abundances for all
other elements that were determined from neutral species were also scaled
by \ion{Fe}{i}. We discuss the basis and validity of this choice in
Section~\ref{sec:discuss}.

In Figure~\ref{fig:FeIdepression} we show the run of $\delta\rm{Fe}$ in the
\citetalias{lapenna16} data.  Apart from the extra scatter added because of
the (small, up to 0.03 dex) differences in \ion{Fe}{II}, this shows the
same trend as the [Na/Fe] differences in Figure~\ref{fig:NaFecompare}. A
linear regression analysis shows the slope is significant ($t = 2.7
\sigma$). The $\delta\rm{Fe}$ values range up to $\sim -0.35$ dex. Also of
note is that there are no stars with an absolute value of $\delta\rm{Fe}$
less than 0.1 dex. The \citetalias{lapenna16} Fe abundances are based on
many lines and have very small reported uncertainties ($\pm 0.01$ dex,
Table 1 of \citetalias{lapenna16}). Thus the entire sample appears to have
highly significant $\delta\rm{Fe}$. \citetalias{lapenna16} conclude that
there is currently no complete explanation of this $\delta\rm{Fe}$ effect
but it ``seems to be a general feature of AGB stars in GCs''. This
conclusion does however rely on the reported uncertainties being
realistic. We address this fundamental condition in
Sections~\ref{sec:realornot} and~\ref{sec:uncerts}.

In summary, we conclude that {\it the differences in [Na/Fe] between
  \citetalias{lapenna16} and \citetalias{campbell13} are driven wholly by
  the \ion{Fe}{I} depression} relative to \ion{Fe}{II} reported by
\citetalias{lapenna16}.  

Our next step in the comparison is to see if
we can reproduce the \citetalias{lapenna16} $\delta\rm{Fe}$ from the
\citetalias{campbell13} FLAMES data.

\section{\ion{Fe}{I} and \ion{Fe}{II} from \citetalias{campbell13} data using \citetalias{campbell13} parameters}\label{sec:FeC13}

   \begin{figure}
   \centering
   \includegraphics[width=0.6\hsize]{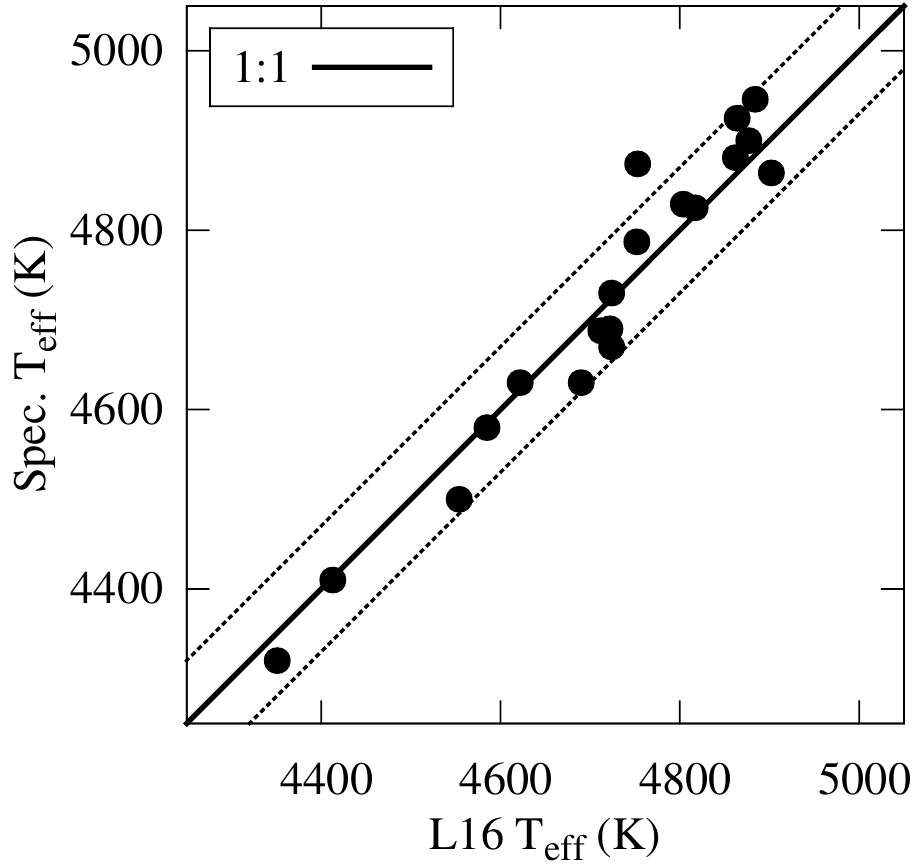}
   \caption{Spectroscopically determined $T_{\rm{eff}}$ for the
     \citetalias{campbell13} data using the L16 method
     (Sec.~\ref{sec:FeC13}), compared to the \citetalias{lapenna16}
     temperatures.  The two sets
     of temperatures were derived using different photometry for initial $T_{\rm{eff}}$
     estimates. Dashed lines indicate typical
     uncertainties ($\pm 70$ K).}
   \label{fig:teff-newstrom}
   \end{figure}

As noted earlier, \citetalias{campbell13} did not derive Fe abundances. A
single Fe abundance was assumed for all stars in their sample, based on a
detailed RGB study ($\rm{[Fe/H]} = -1.54$, \citealt{carretta07}). This was
considered a reasonable assumption since it is well established that NGC
6752 is mono-metallic in Fe
(\citealt{gratton05,carretta09b,yong15}). However it meant that any
unexpected deviation in \ion{Fe}{I} or \ion{Fe}{ii} in the AGB stars would
have been missed.

Here we present newly calculated \ion{Fe}{I} and \ion{Fe}{II}
abundances using the \citetalias{campbell13} FLAMES data, in order to
compare directly with the \citetalias{lapenna16} Fe results.

We derive LTE Fe abundances using the EW method, as in the
\citetalias{lapenna16} study. While \citetalias{campbell13} used the MOOG
stellar line analysis program (\citealt{sneden73}), here we use the WIDTH3
program (\citealt{gratton88a,gratton88b}). We aim to reproduce the
\citetalias{lapenna16} results, so we follow the specific methodology of
that study, which comprises the following steps:
\begin{enumerate}
\item $T_{\rm{eff}}$ is calculated `spectroscopically', i.e. by requiring no
trend between \ion{Fe}{I} abundances and excitation potential.
\item Gravity is adjusted from the initial photometric values by
  recalculating it using the $T_{\rm{eff}}$ from Step~1. Iteration back to Step~1
  may be required if the changes in $\log g$ are significant. Ionisation
  balance is ignored.
\item Microturbulent velocity is then adjusted by requiring no trend between \ion{Fe}{I}
  abundances and line strengths.
\end{enumerate}

We used \cite{kurucz93} model atmospheres, adopting the same $\rm{[M/H]} =
-1.50$ value as \citetalias{lapenna16}. Photometrically-derived values of
$T_{\rm{eff}}$ and $\log g$ were adopted as initial estimates (those in
Fig.~\ref{fig:paramcompare}). The initial temperatures are identical to
those used in \citetalias{campbell13}, based on the Str{\"o}mgren
photometry from \cite{grundahl99} and using the $(b-y)$ relation of
\citet[their eqn. 15]{alonso99}. \object{NGC 6752} suffers from minor
reddening; we corrected the $b$ and $y$ magnitudes for reddening using the
relations of \cite{schlegel98}, adopting $E(B-V)=0.04$~mag
(\citealt{harris96}). The initial $\log g$ values (blue squares in
Fig.~\ref{fig:paramcompare}) were calculated using a stellar mass of 0.61
M$_{\odot}$, and a distance modulus of $(M-m)_{V} = 13.13$
(\citealt{harris96}), consistent with \citetalias{lapenna16}. We used the
bolometric correction relation of \citet[their eqns.~17 and 18]{alonso99}.

Using Step 2 above for gravity estimation one avoids \ion{Fe}{I} being
forced to be equal to \ion{Fe}{II} (i.e. ionisation balance is not
enforced). In iterating back to Step 1 we found that the $\log g$ values
are insensitive to the $\sim 0 \rightarrow 100 \rm{K}$ modifications in
$T_{\rm{eff}}$, with the average change in $\log g$ being $\sim +0.03$
dex. Our initial microturbulence values were estimated using the relation
from \cite{gratton96}. Most of these values were unchanged in Step 3, with
four AGB stars changing by $\sim \pm 0.1$ km/s, so they are still
consistent with those in the right panel of
Figure~\ref{fig:paramcompare}. Our final spectroscopic $T_{\rm{eff}}$
values are consistent with the \citetalias{lapenna16} spectroscopic
temperatures (Fig.~\ref{fig:teff-newstrom}). Finally, our linelist is based
on that of \cite{gratton03}. We explore line list differences in
Sec.~\ref{sec:gfs}.

   \begin{figure}
   \centering
   \includegraphics[width=\hsize]{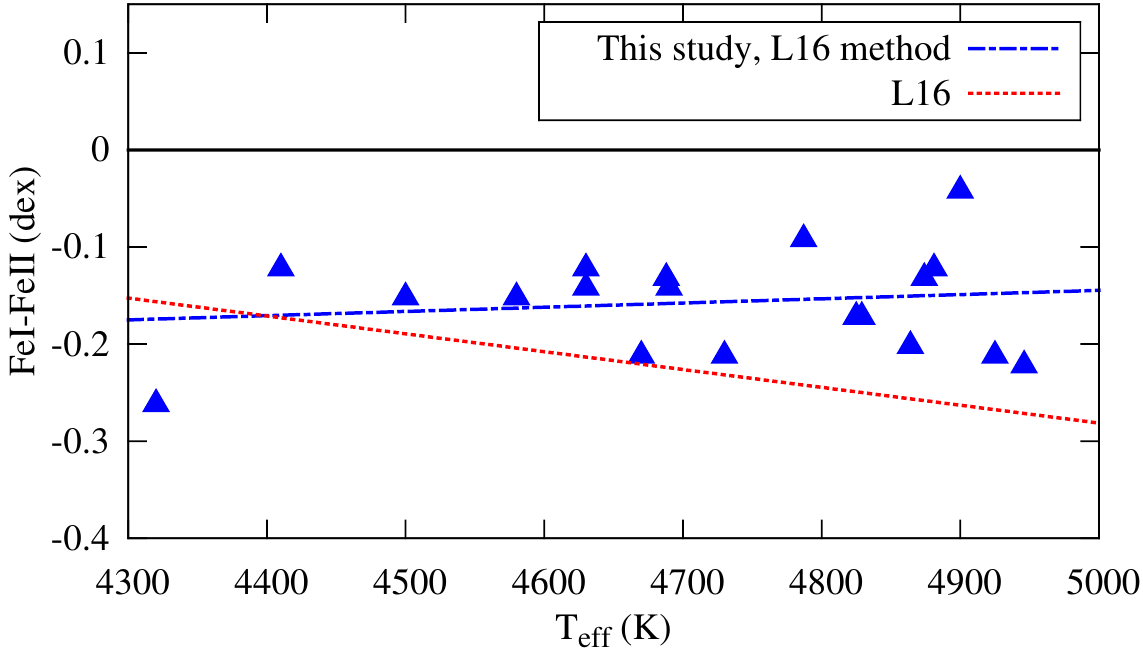}
   \caption{ Depression of \ion{Fe}{I} relative to
     \ion{Fe}{II} we find when using the method of \citetalias{lapenna16} and
     the \citetalias{campbell13} data (blue triangles). The blue dot-dashed line is a
   linear fit to our results, and the dashed red line is the fit to the
   \citetalias{lapenna16} results (from Fig.~\ref{fig:FeIdepression}). The
   temperatures used for this analysis are displayed in Fig.~\ref{fig:teff-newstrom}.}
   \label{fig:C17FeIdepression}
   \end{figure}

With these parameters, and assuming a solar abundance for Fe of $\log
\epsilon = 7.50$ (\citealt{GS98}), we find for the AGB stars:
\begin{itemize}
   \item[] $\rm{[\ion{Fe}{II}/H]_{AGB}} = -1.48
\pm 0.01 dex$ ($\sigma = 0.04$; \citetalias{lapenna16}: $-1.58$ dex)
   \item[] $\rm{[\ion{Fe}{I}/H]_{AGB}} = -1.63 \pm 0.01$ dex ($\sigma = 0.04$;
\citetalias{lapenna16}: $-1.80$ dex). 
\end{itemize}
Thus we confirm a significant \ion{Fe}{I}-\ion{Fe}{II} difference, at least
qualitatively. Unlike the \citetalias{lapenna16} $\delta\rm{Fe}$ results
our results show no substantial trend with $T_{\rm{eff}}$
(Fig.~\ref{fig:C17FeIdepression}), with the significance of the slope being $<
1\sigma$, and $r^{2} = 0.03$. The average value of the offset in our
results is $\delta\rm{Fe} = -0.15 \pm 0.01$ dex ($\sigma = 0.05$), as
compared to $-0.22$ dex in \citetalias{lapenna16}. Using the
\citetalias{campbell13} parameters and the \citetalias{lapenna16} spectra,
\citetalias{lapenna16} found an average offset of $-0.27$ dex. Thus there
is a systematic difference of order 0.1 dex between the studies even
if using the same stellar parameters. This may be related to the 0.1 dex
lower [\ion{Fe}{II}/H] found by \citetalias{lapenna16}, which could be due to
the adoption of different oscillator strengths between the studies (we
explore this as an uncertainty in Sec.~\ref{sec:gfs}).

%--------------------------------------------------------------------
\section{The origin of $\delta\rm{Fe}$}\label{sec:realornot}

The qualitatively similar finding of a definite $\delta\rm{Fe}$ in both
sets of data is in one way reassuring -- it shows that, given a particular
methodology, the results of \citetalias{lapenna16} are reproducible with
independent data and tools.  However, the \citetalias{lapenna16} study
did not investigate the robustness of this result. An obvious question
arises -- is there some systematic problem(s) in the method mimicking this
phenomenon?

To investigate this possibility we explore the uncertainties in the
abundance analysis process.  We begin by noting that it is well known that
(i) offsets in \ion{Fe}{II} can be caused by offsets in gravity, (ii)
offsets in \ion{Fe}{I} can be caused by offsets in $T_{\rm{eff}}$, and (iii) the
magnitude of non-LTE effects is predicted to be small in these stars. We
explore the first two sources of uncertainty in the next two subsections
and the third in Section \ref{sec:nlte}.

\subsection{Gravity check: \ion{Fe}{II} abundance comparison with RGB stars}\label{sec:gravcheck}

One way to check if there is a gravity offset problem is to compare the
\ion{Fe}{II} abundance of the AGB stars to that of the RGB stars -- they
should be identical for \ion{Fe}{II} since NLTE effects are predicted to be
negligible for \ion{Fe}{ii} in late-type stars (eg. \citealt{lind12}). Due
to its dependence on gravity, a difference in \ion{Fe}{II} may indicate
systematic problems with $\log g$ that would require investigation.

\citetalias{campbell13} included RGB stars in their study, as a control
sample. For the Fe determination in the RGB stars we again used the same
methodology of \citetalias{lapenna16}, as described for the AGB sample
above. In this case our results show no evidence of an
\ion{Fe}{i}-\ion{Fe}{ii} discrepancy:
\begin{itemize}
  \item[] $\rm{[\ion{Fe}{II}/H]_{RGB}} = -1.47 \pm 0.01$~dex ($\sigma =
0.06$)
  \item[] $\rm{[\ion{Fe}{I}/H]_{RGB}} = -1.48 \pm 0.04$~dex ($\sigma =
0.06$). 
\end{itemize}
Formally we measure $\delta\rm{Fe} = -0.01 \pm 0.02$ ($\sigma =
0.08$).

Importantly, the AGB $\rm{[\ion{Fe}{II}/H]}$ is perfectly in agreement with
the RGB measurement. This suggests that \ion{Fe}{II} is not the source of
the AGB $\delta\rm{Fe}$ phenomenon. Although not a definitive proof, it
also suggests that the AGB $\log g$ values are reasonable. Assuming this is
correct we now investigate the sensitivity of Fe to $T_{\rm{eff}}$.

\subsection{Temperature check: \ion{Fe}{I} behaviour with varying $T_{\rm{eff}}$}

   \begin{figure}
   \centering
   \includegraphics[width=\hsize]{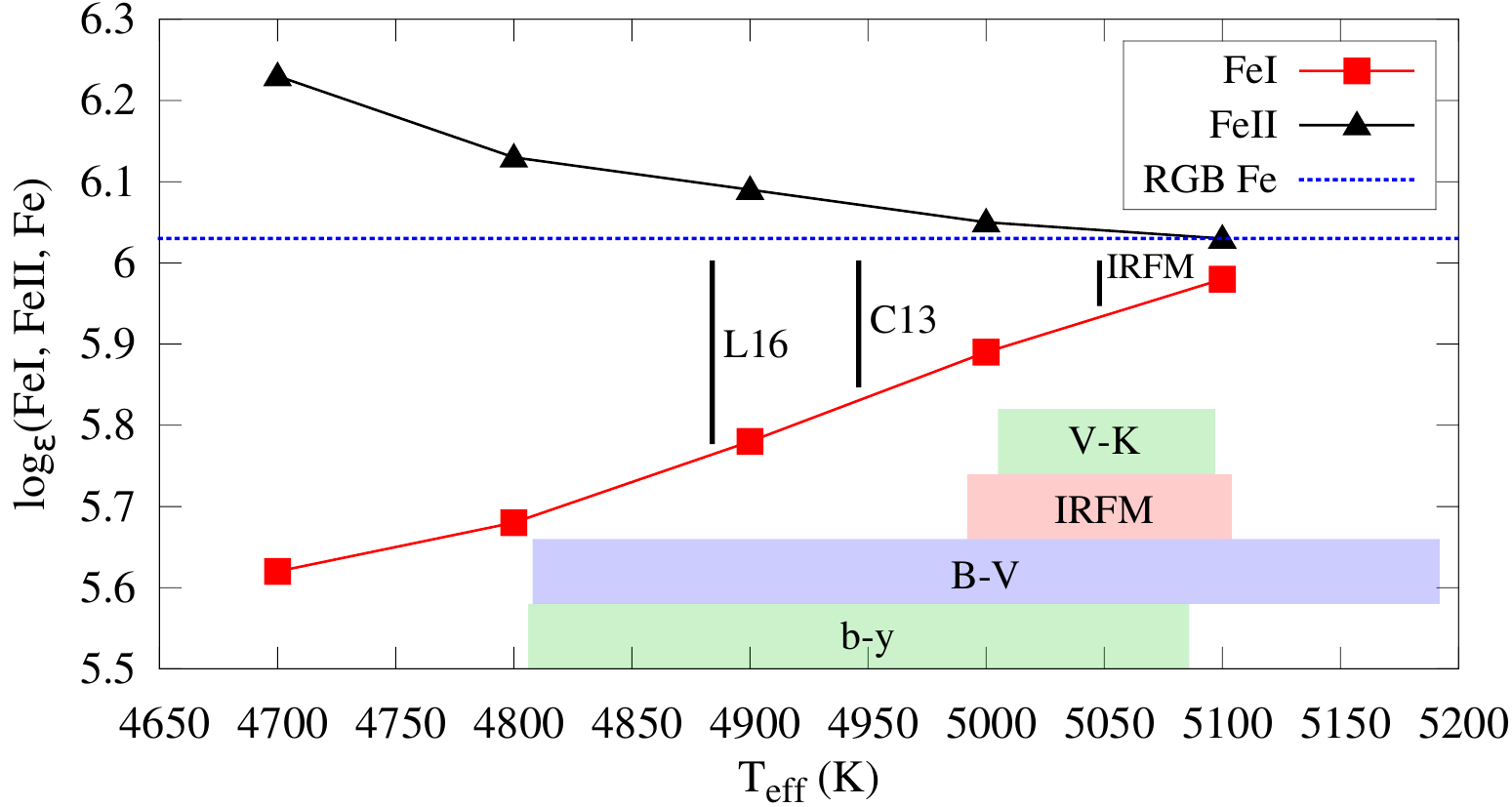}
   \caption{Testing the effect of adopted $T_{\rm{eff}}$ on the derived
     \ion{Fe}{i} and \ion{Fe}{ii} abundances in the AGB star 97. Horizontal
     bars show the $2\sigma$ uncertainty ranges\protect\footnotemark~for
     each of the $T_{\rm{eff}}$-colour relations, centred on the
     $T_{\rm{eff}}$ prediction of each relation (see text for details). The
     temperatures used in the \citetalias{campbell13} study (4946~K) and the
     \citetalias{lapenna16} study (4884~K) are indicated by vertical
     lines, highlighting the magnitude of $\delta\rm{Fe}$ at each
     $T_{\rm{eff}}$. Also shown is our IRFM $T_{\rm{eff}}$ (5048~K) and
     associated uncertainty. The $(V-K)$ $T_{\rm{eff}}$ is 5051~K. The
     dashed line shows the average iron abundance of the RGB stellar
     sample.}
   \label{fig:tefftest}
   \end{figure}

\footnotetext{We use $2\sigma$ uncertainties due to the fact that AGB stars
  are not significantly represented in the stellar samples on which the
  colour-$T_{\rm{eff}}$ relations are based.}

\citetalias{lapenna16} derived surface temperatures spectroscopically, i.e.
by requiring no trend between iron abundances and excitation potential. In
this procedure it is usual to use photometric $T_{\rm{eff}}$ as an initial
estimate. \citetalias{lapenna16} did not specify if this was done, but we
assume it was. We also assume the same BV photometry \citep{stetson00} that
was used for the $\log g$ derivation was also used for
$T_{\rm{eff}}$. Regardless of the source of photometry, and the method to
arrive at the final temperatures, it can be seen in
Figure~\ref{fig:paramcompare} that the \citetalias{lapenna16} temperatures
agree with the \citetalias{campbell13} temperatures, within the
uncertainties. Here we explore the uncertainties, to ascertain whether
systematic problems with $T_{\rm{eff}}$ could be giving rise to the
$\delta\rm{Fe}$ phenomenon present in both studies
(Figs.~\ref{fig:FeIdepression} and \ref{fig:C17FeIdepression}).

\subsubsection{AGB star test case}\label{sec:A97}

To investigate the sensitivity of \ion{Fe}{I} and \ion{Fe}{II} to the
adopted $T_{\rm{eff}}$ we chose one star as a case study: AGB star 97. This
star was chosen because it displays a strong $\delta\rm{Fe}$ signal in both
\citetalias{lapenna16} and the current study, with $\delta\rm{Fe} = -0.31$
and $-0.22$, respectively. In \citetalias{lapenna16} the adopted surface
temperature for this star was 4884~K. In the current study we found the
photometric $T_{\rm{eff}}$ of \citetalias{campbell13} to require no change
(4946~K). The difference of 62~K is within the $1\sigma$ ($\pm 70$~K)
uncertainties of the Str{\"o}mgren relation which we used to derive
$T_{\rm{eff}}$ (\citealt{alonso99}).

For the test we varied $T_{\rm{eff}}$ and attempted to find spectroscopic
`solutions' (i.e. no trend between iron abundances and excitation
potentials) at each $T_{\rm{eff}}$. During this process $\log g$ was kept
constant, at the photometric value. In Section \ref{sec:FeC13} we showed
that the $\log g$ adjustment is negligible within the $T_{\rm{eff}}$
uncertainty ranges considered here.

In Figure~\ref{fig:tefftest} we show the results of the
test. Interestingly, we were able to find spectroscopic `solutions' for a
wide range of temperatures, even outside the uncertainties of the
colour-$T_{\rm{eff}}$ relations, although no solution was found above
5100~K\footnote{Within our test procedure. Varying $\log g$ may allow
    solutions at higher $T_{\rm{eff}}$.}. Multiple solutions were possible
because of the uncertainty in the abundance-excitation potential slope,
combined with the poorly constrained microturbulence parameter, which was
adjusted to reduce the slope in the usual procedure
(Sec. \ref{sec:FeC13}). The slope uncertainty in this case was $\pm
0.03$~dex/eV\footnote{Across the AGB sample the average uncertainty was $\pm
  0.02$~dex/eV.}. Over the the $T_{\rm{eff}}$ range of $4800 \rightarrow
5100$~K the range of microturbulence values we found spanned $1.20
\rightarrow 1.65$~km/s, with the microturbulent velocity increasing with
temperature. Apart from the very low $T_{\rm{eff}}$ end, which is very
unlikely to be representative of the true temperature (Sec.~\ref{sec:VK}),
these appear to be reasonable values, as compared to those reported by
\citetalias{campbell13} and \citetalias{lapenna16}
(Fig.~\ref{fig:paramcompare}). It is also useful to remember that
`microturbulent velocity' is essentially a free parameter, i.e. it has
little physical basis (see eg. the four listed points in Sec.~1 of
\citealt{mucciarelli11}, and references therein). For all solutions there
was no trend between Fe abundances and line strength, within the
uncertainties.

The $\delta\rm{Fe}$ variation over the $T_{\rm{eff}}$ test range shows a
consistent trend: $\delta\rm{Fe}$ decreases with increasing
$T_{\rm{eff}}$. Ignoring the $T_{\rm{eff}}$ values outside the photometric
$T_{\rm{eff}}$ uncertainties, $\delta\rm{Fe}$ ranges from $-0.46$ (at 4820
K) to $-0.05$ (at 5100 K). This final value is consistent with zero given
the $1\sigma$ scatter in $\delta\rm{Fe}$ of 0.08 dex that we found in the
RGB sample (Sec.~\ref{sec:FeC13}).

Also marked in Figure~\ref{fig:tefftest} are the temperatures used by
\citetalias{campbell13} and \citetalias{lapenna16}. Importantly the
$\delta\rm{Fe}$ value at the \citetalias{lapenna16} $T_{\rm{eff}}$ is very
similar to that reported by \citetalias{lapenna16} (their $-0.31$ dex
versus $-0.34$ dex here). Considering that different spectra and tools have
been used, and that both the \citetalias{lapenna16} and
\citetalias{campbell13} $\delta\rm{Fe}$ values fit the
$\delta\rm{Fe}$-$T_{\rm{eff}}$ trend, this is a strong confirmation of the
$\delta\rm{Fe}$ phenomenon, and its dependence on $T_{\rm{eff}}$, both
qualitatively and quantitatively.

The gradient $\delta\rm{Fe}$/$T_{\rm{eff}}$ is $\sim 0.002$ dex/K. Given
a typical $1\sigma$ $T_{\rm{eff}}$ uncertainty of $\pm 100$K for the $(B-V)$
colour, this translates to a possible $\delta\rm{Fe}$ range of 0.40
dex. This is a very substantial uncertainty and consistent with the up to
0.35 dex found by \citetalias{lapenna16}.

\subsubsection{Ramifications of the $\delta\rm{Fe}$ dependence on $T_{\rm{eff}}$}\label{sec:ramifications}

This result shows clearly that significant $\delta$Fe values can arise even
within the photometric $T_{\rm{eff}}$ uncertainties.  Crucially it appears
that the initial $T_{\rm{eff}}$ estimate (usually photometric) is central
in determining the final $\delta\rm{Fe}$ value. This is because there is a
continuum of spectroscopic `solutions' across the $T_{\rm{eff}}$
uncertainty range, so that {\it the spectroscopically determined
  $T_{\rm{eff}}$ will usually be close to the photometric estimate}. Figure
\ref{fig:tefftest} then implies that, if there is a systematic trend or
offset in the inferred photometric temperatures, a similar trend or offset
should be present in $\delta\rm{Fe}$ -- even if the temperatures are
determined spectroscopically. Given this, an investigation into the
sources of the adopted temperatures is mandatory, and is our next step.

\subsubsection{Temperature scales and the case for $(V-K)$}\label{sec:VK}

In Figure~\ref{fig:tefftest} we also show the predictions of three
colour-$T_{\rm{eff}}$ relations for our AGB test star: Str{\"o}mgren $(b-y)$
(\citealt{alonso99} eqn.~15, with a quoted uncertainty of $1\sigma =
70$~K), Johnson $(B-V)$ (\citealt{alonso99} eqn.~4, $\sigma = 96$~K), and
Johnson-2MASS $(V-K_{\rm{s}})$ (Table 5 of \citealt{gonzalezhernandez09}, $\sigma
= 23$~K). Reddening was corrected for in $(V-K_{\rm{s}})$ using the
relation of \citet[their Eqn. 8]{fitzpatrick07} assuming $R_{V} = 3.1$ and
$E(B-V) = 0.04$.

As a cross-check we have also calculated our own $T_{\rm{eff}}$ for this
star using the \cite{casagrande10} implementation of the infrared flux
method (IRFM). The IRFM estimates $T_{\rm{eff}}$ by comparing the ratio of
the observed bolometric flux to a monochromatic IR flux with the ratio
predicted by theory (synthetic spectra). Since the synthetic spectra have a
very mild dependency on stellar parameters in the IR, this method is only
weakly dependent on the models. The \cite{casagrande10} scale is calibrated
absolutely, using a set of solar twins.  For further details of our IRFM
procedure we refer the reader to \cite{casagrande10}. We used the 2MASS JHK
(\citealt{skrutskie06}) and BV photometry for this $T_{\rm{eff}}$
determination (and for all the IRFM temperatures in this study). The
temperature we derived has an internal uncertainty of $\pm30$~K, and is
also included in Figure~\ref{fig:tefftest}. The BV photometry we use in
this study is from \cite{momany02}. These data are of high quality, for
example the average error on the V magnitudes for the AGB sample is 0.008
mag.

Immediately obvious from Figure~\ref{fig:tefftest} is that the $(V-K)$
relation gives the most precise $T_{\rm{eff}}$ estimate. It is also in
perfect agreement with our IRFM-derived $T_{\rm{eff}}$, which has a similar
degree of precision. Interestingly, both of these $T_{\rm{eff}}$ estimates
give much lower $\delta\rm{Fe}$ values than obtained using either the
\citetalias{campbell13} or \citetalias{lapenna16} temperatures, with
$\delta\rm{Fe}$ approaching zero at the higher end of the $2\sigma$
uncertainty bands.

That $(V-K)$ has a small uncertainty for late-type stars is well known and
is due to it being (i) only marginally sensitive to metallicity/line
blanketing, and (ii) having a negligible dependence on surface gravity
(\citealt{alonso99,ramirez05}). Importantly for our study, the
$(V-K)$ colour is particularly suited to giants. Indeed \cite{alonso99}
suggest that it is ``probably the best temperature indicator for giant
stars''.  Furthermore, \cite{ramirez05} report that, due to the colour
being so insensitive to gravity, particularly in the range 4800~K~$>
T_{\rm{eff}} > 6000$~K, it makes $(V-K)$ suitable for stars of unknown
luminosity class. This is important for studies of (early) AGB stars
because many of them lie in this $T_{\rm{eff}}$ range (our sample: $4500
\rightarrow 5050$~K) and it is a class of stars that have only recently
started to be investigated in detail, so their surface gravities are less
certain than RGB star gravities.

   \begin{figure}
   \centering
   \includegraphics[width=\hsize]{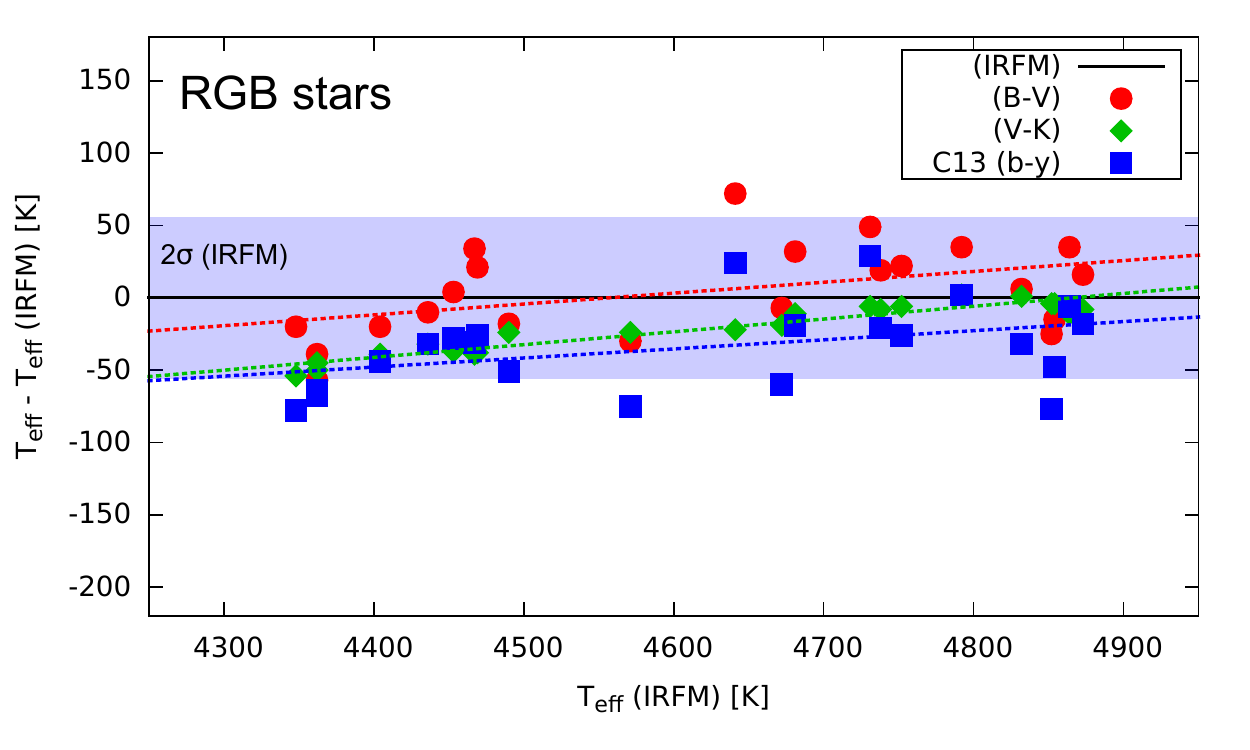}
   \caption{Comparison of $T_{\rm{eff}}$ derived for the RGB stars using four
     different methods: three different colour-$T_{\rm{eff}}$ relations and
     our IRFM (see text for details). The $T_{\rm{eff}}$ from the
     colour-$T_{\rm{eff}}$ relations are shown as differences from the IRFM
     temperatures. Dashed lines are linear fits to each $T_{\rm{eff}}$ set,
     and the shaded area shows the average $2\sigma$ internal uncertainty of the
     IRFM scale ($\pm 56$ K). The $2\sigma$ uncertainties of the $(b-y)$ and
     $(B-V)$ relations are much larger than for our IRFM, being $\pm140$ and
     $\pm 192$ K, respectively \citep{alonso99}. The $(V-K)$ uncertainties
     ($2\sigma = 46$ K; \citealt{gonzalezhernandez09}) are similar to the
     IRFM uncertainties.\label{fig:teffsRGB}}
   \includegraphics[width=\hsize]{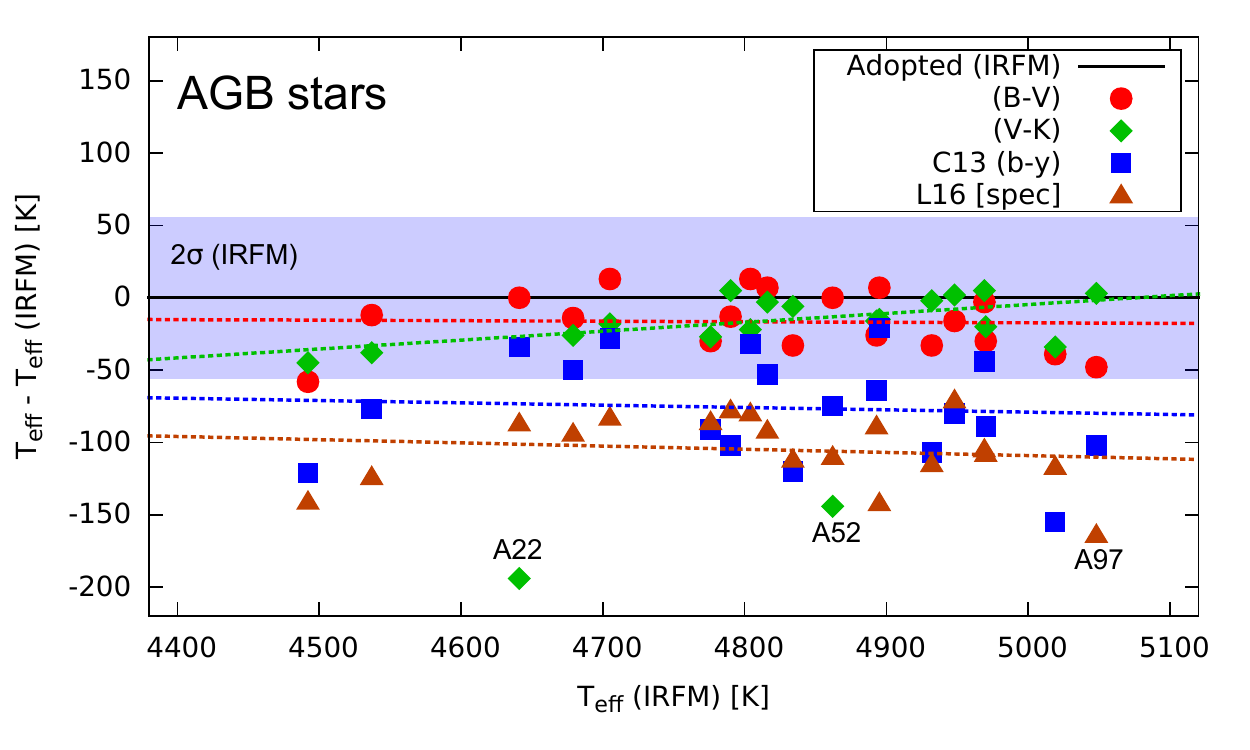}
   \caption{Comparison of five $T_{\rm{eff}}$ determinations for the AGB star
     sample. Symbols and shading are the same as Fig.~\ref{fig:teffsRGB},
     except for the addition of the \citetalias{lapenna16} temperatures
     (\citetalias{lapenna16} did not study RGB stars). Three stars are
     highlighted with labels: AGB star 97 was the subject of our
     $\delta\rm{Fe}$ tests (Fig.~\ref{fig:tefftest}), whilst A22 and
     A52 have uncertain IRFM and $(V-K)$ $T_{\rm{eff}}$ due to suspect
     2MASS~K~magnitudes. These latter two stars have not been included in
     the linear regression lines for $(V-K)$ or IRFM, and we adopt the $(B-V)$
     temperatures for them.\label{fig:teffsAGB}}
   \end{figure}

\subsubsection{Temperature scales: Ensemble comparisons} 

As a further check of the $T_{\rm{eff}}$ `scales' we now perform ensemble
comparisons between the $T_{\rm{eff}}$ predictions from the same three
colour-$T_{\rm{eff}}$ relations detailed above but across our entire
\object{NGC~6752} RGB and AGB samples. We also present our IRFM
temperatures for all the stars.

The RGB sample comparison is displayed in
Figure~\ref{fig:teffsRGB}. Although small offsets and small temperature
trends are present, it is clear that all three relations give temperatures
that are consistent with each other, within the quoted $1\sigma$
uncertainties \citep{alonso99,gonzalezhernandez09}. This is true across the
whole $T_{\rm{eff}}$ range of our RGB sample. This confirms the
well-constrained nature of the parameters for GC RGB stars, as expected
from much previous work on these types of stars.

The AGB on the other hand has not been well studied. In
Figure~\ref{fig:teffsAGB} we show $T_{\rm{eff}}$ for the AGB
stars. Here the $(B-V)$, $(V-K)$, and IRFM temperatures are consistent with
each other, similar to the RGB case. However the Str{\"o}mgren $(b-y)$
temperatures (used by \citetalias{campbell13}) are offset by about $-60$
K. This is particularly true at higher $T_{\rm{eff}}$ ($> 4700$~K), where
the majority of the temperatures are outside the $2\sigma$ uncertainties of
the IRFM $T_{\rm{eff}}$. Also displayed are the temperatures from
\citetalias{lapenna16}. These are offset even more, by about $-100$~K on
average.

Given that we have showed in Figure~\ref{fig:tefftest} that $\delta\rm{Fe}$
is is strongly correlated with a reduction in $T_{\rm{eff}}$, this is very
suggestive that the large $\delta\rm{Fe}$ reported by
\citetalias{lapenna16} (Fig. \ref{fig:FeIdepression}) is driven by the
$T_{\rm{eff}}$ scale of that study. It also explains our own finding of
significant $\delta\rm{Fe}$ using the Str{\"o}mgren $(b-y)$ temperatures
adopted by \citetalias{campbell13}. That the temperature scale of
\citetalias{campbell13} is slightly warmer than that of
\citetalias{lapenna16} also shows why our $\delta\rm{Fe}$ values are
generally smaller in magnitude than those of \citetalias{lapenna16}
(Sec.~\ref{sec:FeC13}; Fig.~\ref{fig:C17FeIdepression}).

The next logical step is to use the more appropriate temperatures in
deriving Fe abundances. The change in $T_{\rm{eff}}$ scale may also affect
$\ion{Na}{I}$, which we also re-derive in Section~\ref{sec:newsodium}.

%--------------------------------------------------------------------
\section{Iron from \citetalias{campbell13} data using new $T_{\rm{eff}}$ scale}

\subsection{Reanalysis method and results}

\begin{figure}
  \centering
  \includegraphics[width=0.8\hsize]{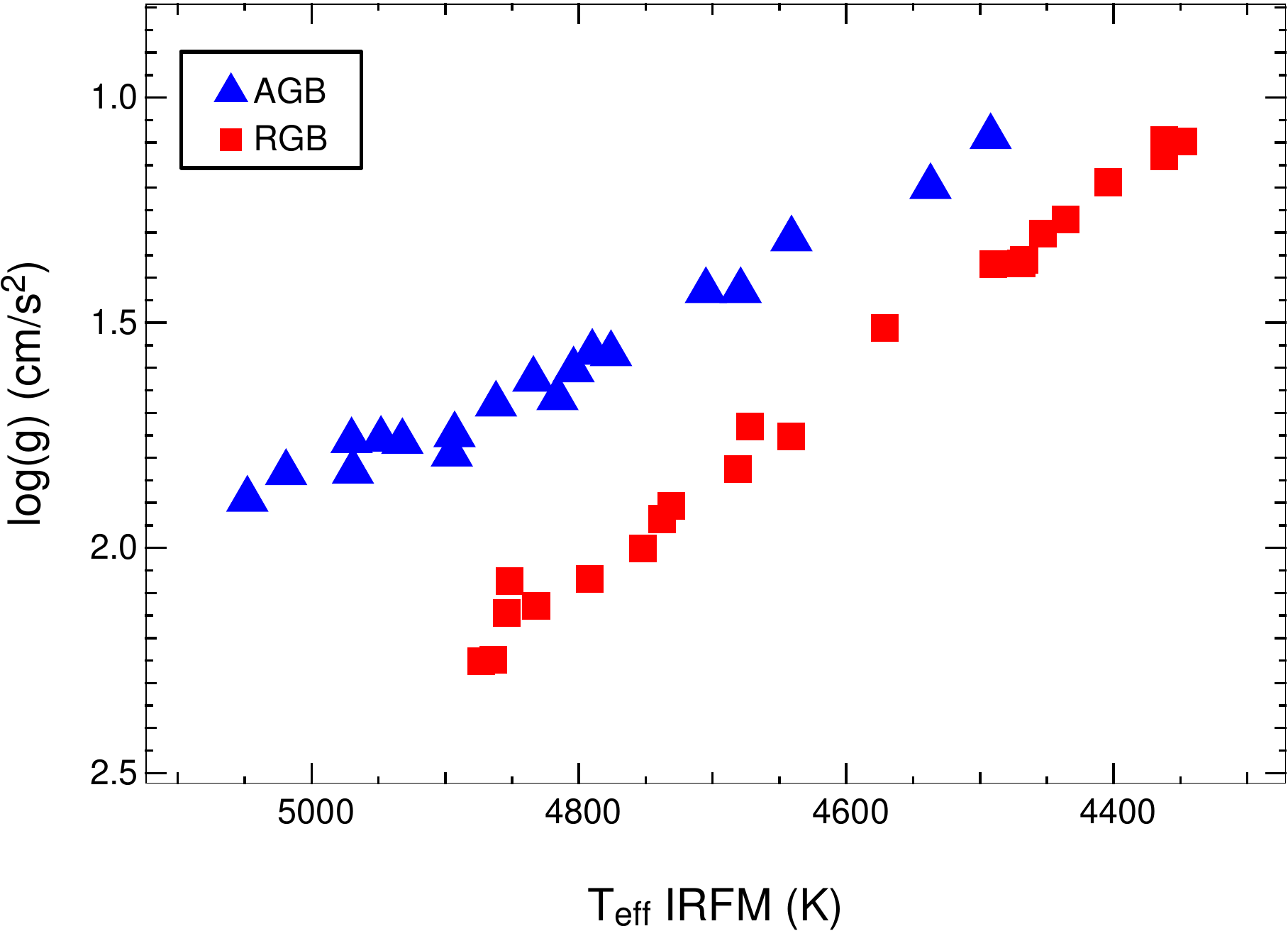}
  \caption{Our final stellar parameters for the \object{NGC 6752} AGB and
    RGB stars. All are `photometric' -- based on the IRFM temperatures
    calculated for this study, except for 2 AGB stars (see
    Fig.~\ref{fig:teffsAGB}).}
  \label{fig:cmd}
\end{figure}

In our final reanalysis of the \citetalias{campbell13} spectra we chose to
use photometric parameters (IRFM) only, because (i) the temperature scale
appears quite accurate so we want to avoid additional uncertainties by
using the spectroscopic $T_{\rm{eff}}$ method, and (ii) following
\citetalias{lapenna16}, we do not want to
force \ion{Fe}{I} = \ion{Fe}{II} by obtaining $\log g$
spectroscopically. We adopt the same distance modulus as
\citetalias{campbell13} and \citetalias{lapenna16}, a mass of
0.78~M$_{\odot}$ for RGB stars and 0.61~M$_{\odot}$ for AGB
stars. Microturbulent velocities were estimated using the \cite{gratton96}
relation. Our final stellar parameters are plotted in the $\log
g$-$T_{\rm{eff}}$ plane in Figure \ref{fig:cmd}, and listed in
Table~\ref{tab:bigtab}. Note that there are 19 AGB stars rather than 20,
since for one star we did not have all three sets of photometry (star 89 of
\citetalias{campbell13} and \citetalias{lapenna16}).

\begin{figure}
  \centering
  \includegraphics[width=\hsize]{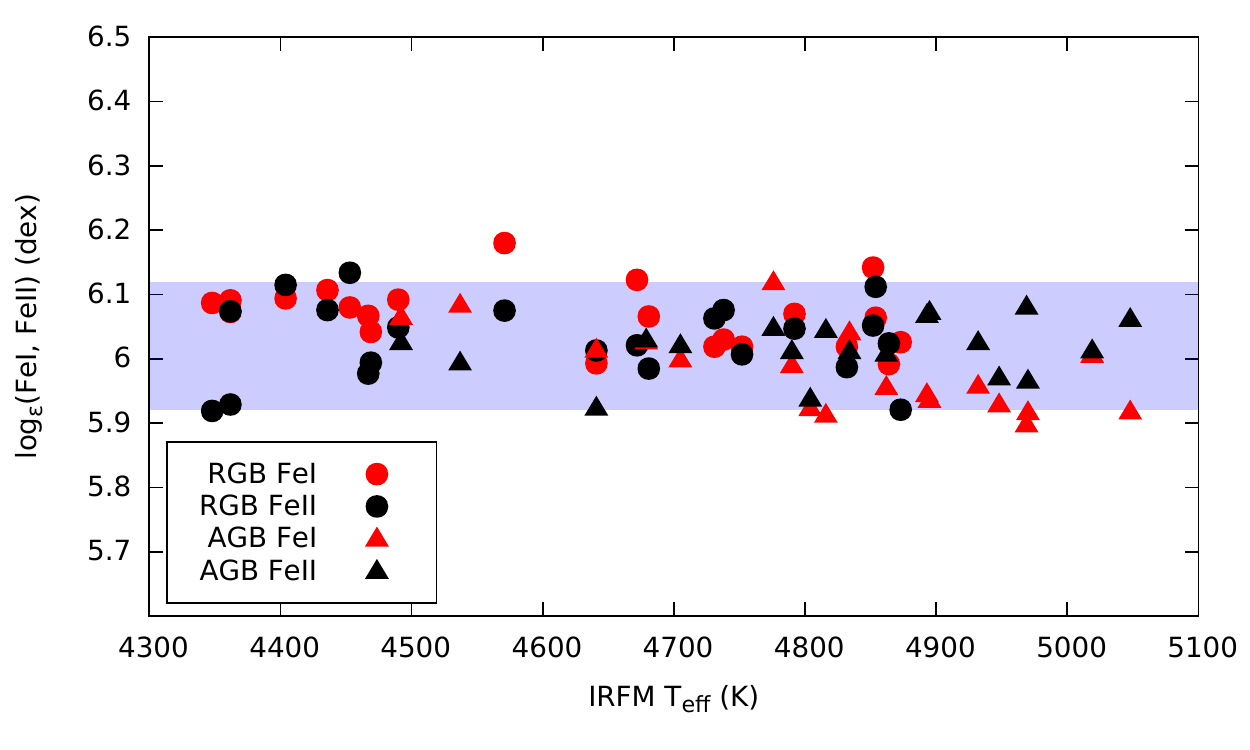}
  \caption{Iron results using our new IRFM temperatures. All RGB stars
    (dots) and AGB stars (triangles) are shown for comparison, with
    iron abundances derived using \ion{Fe}{I} and \ion{Fe}{II} highlighted
    by colour (red and black respectively). The shaded region indicates a
    typical uncertainty of $\sim \pm 0.10$ dex, centered around the average
    \ion{Fe}{II} abundance ($\log{\epsilon} = 6.02$, from RGB and AGB
    stars).  \label{fig:newFe}}
  \includegraphics[width=\hsize]{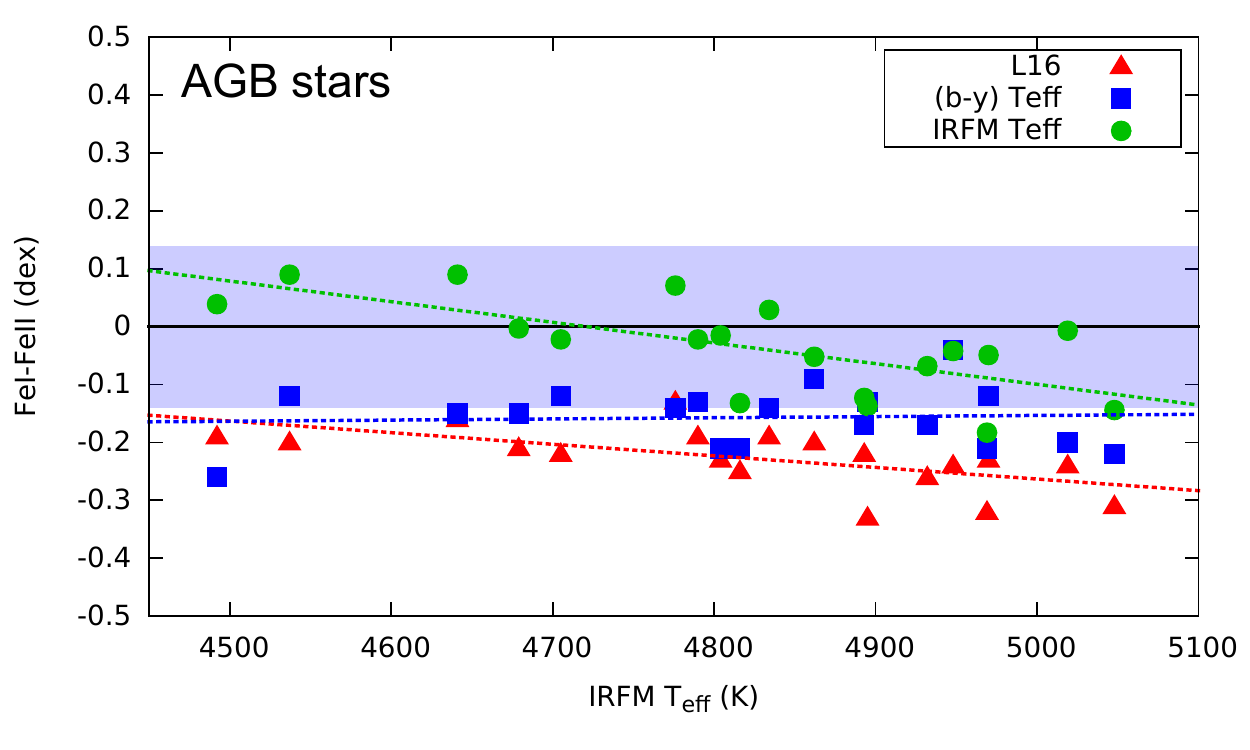}
  \caption{AGB $\delta\rm{Fe} = \rm{\ion{Fe}{I}}-\rm{\ion{Fe}{II}}$ results for
    the current study obtained using our IRFM temperatures, and the $(b-y)$
    temperatures. Also shown are the \citetalias{lapenna16} values. The
    shaded region indicates a quadratic sum of typical ($1\sigma$) \ion{Fe}{I} and
    \ion{Fe}{II} uncertainties ($\pm 0.14$ dex). \label{fig:newdFecompare}}
\end{figure}

In Figure \ref{fig:newFe} we show the final iron results for our whole
sample of RGB and AGB stars. \ion{Fe}{I} and \ion{Fe}{II} are shown
separately for each star. Immediately obvious in this figure is that all
abundances fall within the expected uncertainty range characterised by
$1\sigma \sim 0.1$~dex. Final Fe abundances are also listed in
Table~\ref{tab:bigtab}.

The main effect of the new stellar temperatures is to raise the \ion{Fe}{I}
values in the AGB sample, as expected from Figures~\ref{fig:tefftest},
\ref{fig:teffsRGB}, and \ref{fig:teffsAGB}. Table~\ref{tab:finalfe} shows
that the average increase in \ion{Fe}{I} is $+0.11$ dex, as compared to our
results using the \citetalias{campbell13} stellar parameters
(Sec.~\ref{sec:FeC13}). \ion{Fe}{II} is unchanged, so this translates
directly into a reduction of average $\delta\rm{Fe}$, reducing it from
$-0.15$ to $-0.04$ dex. Figure \ref{fig:newdFecompare} shows visually that
$\delta\rm{Fe}$ in the AGB sample is now negligible. A weak trend appears
to be visible though, with $\delta\rm{Fe}$ increasing in magnitude in the
hotter stars ($T_{\rm{eff}} > 4800$~K). The average $\delta\rm{Fe}$ is
however only about $-0.1$~dex in this subset of AGB stars, and the trend
mostly lies within the error band. We speculate that this possible trend
may be due to either residual underestimation of $T_{\rm{eff}}$, or due to
NLTE effects being stronger in the hotter AGB stars (although the latter is
not supported by current theory, see Sec.~\ref{sec:nlte}).

The main feature in the average \ion{Fe}{I} and \ion{Fe}{II} values
(Table~\ref{tab:finalfe}, final set) is that \ion{Fe}{I} is lower in the
AGB stars than in the RGB stars, by $0.11$~dex. This difference is just
within the combined $ 1\sigma$ dispersions of each sample (0.05 and 0.06,
respectively; Table~\ref{tab:finalfe}), so it is marginally
significant. The difference becomes even less significant when considering
other uncertainties (Sec.~\ref{sec:uncerts}).

Averaging Fe abundances from \ion{Fe}{I} and \ion{Fe}{II} in the AGB and
RGB samples to arrive at total [Fe/H] values shows that the difference
between the two evolutionary phases is $-0.06$ dex, which is comparable to
the star-to-star scatter. Visually the closeness of all the Fe
determinations can be seen in Figure \ref{fig:newFe}.

We also computed abundances for the AGB stars using the $(V-K)$
temperatures. We found this temperature scale to give identical
$\delta\rm{Fe}$ to the IRFM scale. This was expected since the temperatures
are very similar, as seen in Fig. \ref{fig:teffsAGB}. We now explore other
uncertainties in the method.

\begin{table*}[]
\centering
\caption{Summary of iron abundances derived from \ion{Fe}{I} \&
  \ion{Fe}{II} using different input parameters. }
\label{tab:finalfe}
\begin{tabular}{l|cc|cc|cc}
\hline\hline
Analysis run     & [\ion{Fe}{I}/H]  & $\sigma$ & [\ion{Fe}{II}/H] & $\sigma$ & $\delta\rm{Fe}$  & $\sigma$ \\
                 & (dex)        &(dex)     &(dex)         & (dex)    &
(dex)   & (dex)\\
\hline
AGB L16                & $-1.80$ & $0.05$ & $-1.58$ & $0.02$ & $-0.22$ & $0.05$ \\
\hline
RGB (C13 $T_{\rm{eff}}$) & $-1.48$ & $0.06$ & $-1.47$ & $0.06$ & $-0.01$ & $0.08$ \\
AGB (C13 $T_{\rm{eff}}$) & $-1.63$ & $0.04$ & $-1.48$ & $0.04$ & $-0.15$ & $0.05$\\
\hline
RGB (L16 gfs)  & $-1.43$ & $0.05$  & $-1.52$  & $0.05$  & $+0.09$ & $0.07$ \\
AGB (L16 gfs)  & $-1.52$ & $0.06$  & $-1.53$  & $0.04$  & $+0.01$ & $0.08$ \\
\hline
RGB (Final)    & $-1.43$ & $0.05$  & $-1.47$  & $0.06$  & $+0.04$ & $0.07$\\  
AGB (Final)    & $-1.52$ & $0.06$  & $-1.48$  & $0.04$  & $-0.04$ & $0.07$ \\
\hline
\end{tabular}
\tablefoot{Abundances are scaled based on a solar Fe
  abundance of $\log\epsilon = 7.50$ dex. Also shown is the difference
  $\delta\rm{Fe} = \rm{\ion{Fe}{I}}-\rm{\ion{Fe}{II}}$. The first line
  shows the \citetalias{lapenna16} AGB results. The second set of results
  were obtained using the \citetalias{campbell13} temperatures
  (Sec. \ref{sec:FeC13}). The 3rd set of results were obtained using our
  new IRFM $T_{\rm{eff}}$ scale but adopting the \citetalias{lapenna16}
  $\log gf$ values. Our final results, using the IRFM temperatures and our
  $\log gf$ values, are in the last two rows. The typical number of 
   \ion{Fe}{I} lines analysed was 20-40, and 2-3 for \ion{Fe}{II}. The $\sigma$ values
  are the $1\sigma$ star-to-star scatter only.}
\end{table*}

\subsection{Sensitivity of Fe to other uncertainties}\label{sec:uncerts}

\subsubsection{Weighted oscillator strengths}\label{sec:gfs}
The weighted oscillator strength ($\log gf$)\footnote{Where $f$ is the
  oscillator strength and $g$ is the statistical weight of the lower level.}
adopted for each line is a known source of uncertainty in spectroscopic
abundance determination (eg. \citealt{gray05}). Since oscillator strength
quantifies the transition probability of a species from one level to the
next, a change in $\log gf$ has a systematic effect on derived abundances,
shifting them to higher or lower values. This is another possible source of
difference between our study and \citetalias{lapenna16} that could directly
affect $\delta\rm{Fe}$, and thus it requires investigation.

As a first step we directly compared our \ion{Fe}{i} and \ion{Fe}{ii} $\log
gf$ values with those used by
\citetalias{lapenna16}\footnote{\citetalias{lapenna16} used the
  Kurucz/Castelli line list for all species except for \ion{Fe}{II}, for
  which they used values from \cite{Melendez09}}.  For \ion{Fe}{i} we found
the average difference for our 40 lines to be $\Delta\log gf = -0.02$~dex
($\sigma = 0.09$; in the sense \citetalias{lapenna16} -- this study). Since
final abundances are taken as an average over the abundances inferred from
each line, and considering other uncertainties, this difference is
insignificant\footnote{We also compared our \ion{Fe}{i} $\log gf$ values to
  those used by the Gaia-ESO Survey (\citealt{ruffoni14}). Here also the
  average difference is minor, with $\Delta \log gf = 0.02$~dex ($\sigma = 0.09$). There
  are however only five lines in common (out of 40).}. For the \ion{Fe}{ii} lines the
average difference is slightly larger, being $\Delta\log gf =
+0.05$~dex. However, for \ion{Fe}{ii} we only used two or three lines, so
in the cases where only two lines were available, even one significantly
deviant $\log gf$ value would be expected to alter the derived abundances
tangibly -- and thus alter $\delta\rm{Fe}$ by offsetting \ion{Fe}{ii}.  To
check the sensitivity of our results to this difference we re-derived Fe
abundances for our entire stellar sample using the \citetalias{lapenna16}
$\log gf$ values for \ion{Fe}{ii}. The \ion{Fe}{ii} lines ($\Delta\log gf$)
we used were: 6149.23~\AA~($+0.04$~dex), 6247.56~\AA~($+0.02$~dex), and
6369.46~\AA~($+0.10$~dex). All else was kept constant.

The results of this analysis are presented in Table \ref{tab:finalfe}. The
slightly increased \ion{Fe}{ii} oscillator strengths led to an average
decrease of [\ion{Fe}{II}/H] of 0.05~dex in both the RGB and AGB
samples. As \ion{Fe}{i} is unchanged, this leads to correspondingly minor
changes in $\delta\rm{Fe}$.  For the RGB sample $\delta\rm{Fe}$ becomes
marginally significant ($+0.09$~dex), albeit in the opposite sense to the
original problem in the AGB sample. In the AGB stars $\delta\rm{Fe}$
remains insignificant, with $\delta\rm{Fe}$ changing from $-0.04$ to
$+0.01$~($\sigma = 0.08$).

Thus the adopted $\log gf$ values appear to contribute to the uncertainty
in $\delta\rm{Fe}$ only to a small degree. Additionally, the lower
abundance derived from \ion{Fe}{ii} using the \citetalias{lapenna16}
oscillator strengths most likely explains part of the $\sim0.1$ dex lower
average [\ion{Fe}{ii}/H] value found by \citetalias{lapenna16}.

\subsubsection{NLTE effects}\label{sec:nlte} 
Our final $\delta\rm{Fe}$ results appear to show a weak trend toward higher
values at higher $T_{\rm{eff}}$ in the AGB sample
(Fig. \ref{fig:newdFecompare}). This could be due to a real overionisation
of \ion{Fe}{I}, but the overionisation effect would have to be stronger in stars
in this particular temperature range ($T_{\rm{eff}} > 4800$~K).

The magnitude of non-LTE effects in atmospheres of cool stars has been
recently studied by \cite{lind12} and \cite{mashonkina16}. Figure 2 of
\cite{lind12} shows that, at the metallicity of NGC 6752, the NLTE
corrections for \ion{Fe}{i} are expected to be small for AGB and RGB
stars. 

As a check we computed the expected NLTE corrections for a range of
\ion{Fe}{i} lines at some characteristic parameters of our AGB and RGB
stars using both the \cite{lind12} and \cite{mashonkina16} web-based
interpolation routines\footnote{The INSPECT interface:
  \url{http://www.inspect-stars.com} (\citealt{lind11}), and the interface by
  \cite{mashonkina16}: \url{http://spectrum.inasan.ru/nLTE}}.  We found that
corrections were consistent between the two compilations. The corrections
were also almost constant, varying by just $\pm 0.01$ dex, so they are
basically offsets. The constancy across the AGB temperature range implies
that the marginal $\delta\rm{Fe}$ trend in Figure~\ref{fig:newdFecompare}
is not explained by current NLTE theory.  The magnitude of the corrections
are however slightly different across each set of stars, with
$\Delta_{NLTE}$ averaging $+0.05$~dex for the RGB stars and $+0.09$~dex for
the AGB stars. The slightly higher value for AGB stars is expected due to
their higher temperatures\footnote {The lower gravity of the AGB stars
  compared to RGB stars at the same temperature reduces the difference
  marginally.}.

The NLTE offsets increase the average $\delta\rm{Fe}$ to $+0.09$ in the RGB
sample, and to $+0.05$ (from $-0.04$) in the AGB sample. Considering other
uncertainties, these offsets are small. It is important to recognise that
the NLTE corrections themselves also have uncertainties.  \cite{lind11}
showed that model atmosphere choice alone can alter the predicted NLTE
corrections by up to $\sim0.1$~dex (their Figure 8, for \ion{Na}{i}). This
is comparable to the magnitude of the predicted offsets we have reported
here. To be consistent with \citetalias{lapenna16}, and considering the
small effect on the results, we did not apply the NLTE offsets to our final
\ion{Fe}{i} results. This also avoids adding in the extra uncertainty of
the corrections themselves.

\subsubsection{Model Atmospheres} 
The choice of model atmosphere has an effect on abundance
determinations. This is due to the fact that different physical
stratifications are predicted by different stellar atmosphere codes -- for
the same set of stellar parameters. It is the differences in adopted
physical descriptions in each set of theoretical models that gives rise to
the different stratifications. For example, some use `pure MLT' to describe
convection, whilst some use modified MLT formalisms. Overshoot (see
eg. \citealt{castelli97}) and the adopted treatment of opacity are other
model variables.

We ran some tests to gauge the effect of using different model grids on the
derived \ion{Fe}{i} and \ion{Fe}{ii} abundances. For the tests we used four
different grids: a MARCS grid\footnote{Downloaded from
  \url{http://marcs.astro.uu.se}} (\citealt{gustafsson08}), and three different
Kurucz/Castelli grids\footnote{Downloaded from
  \url{http://kurucz.harvard.edu/grids.html}} (their NOVER, OVER, and AODFNEW
grids; \citealt{kurucz93,castelli04}). The four grids differ in terms of
overshoot (or lack of), treatment of convection, and treatment of opacity,
for example. The parameters and EWs of one RGB star (star 12) and one AGB
star (star 97) were used.

We found differences of 0.04 to 0.12 dex in the derived \ion{Fe}{i} and
\ion{Fe}{ii} abundances, with \ion{Fe}{i} consistently at the upper
end. This is consistent with the uncertainty due to adopted model
atmospheres reported by \cite{lind11} in relation to NLTE
corrections. This $\sim 0.1$ dex uncertainty is especially important when
considering species that are very temperature sensitive, here \ion{Fe}{i},
since the temperature stratification changes significantly. This test also
indicates that uncertainties of this order must be allowed for when
comparing between independent studies, even if they are based on the same
data, since the model grid choice affects the results. Variations in
tools/pipelines that make use of the model grids must also add to these
uncertainties.

Two other possible sources of uncertainty from model atmospheres are (i)
the choice of plane parallel or spherical (but still 1D) models, and (ii)
the choice of model stellar mass. Traditionally, 1~M$_{\odot}$ stellar
atmospheres are used for GC stars, since there is negligible effect in
changing the mass by small amounts. However, due to the particularly low
masses of AGB stars ($\sim 0.6$~M$_{\odot}$), models with mass of
0.5~M$_{\odot}$ may be more appropriate. Due to their low envelope mass,
the atmospheres of these stars are expected to be more extended than those
of RGB stars, and thus spherical effects may be important. To check these
two factors we made a test using the MARCS 0.5 M$_{\odot}$ spherical
models, for which only a small grid exists (\citealt{gustafsson08}). We
compared the \ion{Fe}{i} and \ion{Fe}{ii} abundances derived using MARCS
models with mass of 1 M$_{\odot}$ with those derived using 0.5 M$_{\odot}$
models, for a star with characteristic AGB parameters. We found that the
differences were negligible, being of the order 0.01~dex. This indicates
that mass and sphericity are not important in the case of these AGB stars.

Finally we note that the discussion above has only involved 1D model
atmospheres. Three dimensional model atmospheres are now becoming available
and have been shown to have significantly different stratifications
as compared to 1D models (eg. \citealt{magic13}). Thus the use of 3D model
atmospheres would be expected to introduce further differences in
abundance determinations.

\subsubsection{Distance} 
The cluster distance is a fundamental parameter that has a direct impact on
the gravity scale through the derivation of the stellar bolometric
magnitudes. All of the studies (\citetalias{campbell13},
\citetalias{lapenna16} and this study) used the \cite{harris96} catalogue
value of $(m-M)_{V} = 13.13$. A literature search showed that this is at
the lower end of the values published, which range from 13.13 to 13.38
(just in the studies we consulted, the range may be greater;
\citealt{renzini96,gratton03,yong05}). Taking the maximum of these values
systematically decreases gravity by 0.1~dex. This uncertainty in $\log g$
translates to a systematic shift of the microturbulent velocity scale by
an insignificant amount ($+0.03$ dex, based on the \citealt{gratton96}
relation). Nevertheless, we tested the effect of these small systematic
shifts using the parameters of AGB star 97 and RGB star 12. As expected,
\ion{Fe}{i} was unaffected and \ion{Fe}{ii} was reduced by $\sim
0.05$~dex. This reduced $\delta\rm{Fe}$ in the AGB star, from $-0.14$ to
$-0.10$~dex, and increased it in the RGB star since \ion{Fe}{i} was already
greater than \ion{Fe}{ii}. Again these are small changes but they do add to
the many other small uncertainties.

\subsubsection{Effect of AGB stellar mass on gravity} 
Another uncertainty affecting gravity determination is the adopted stellar
mass for the AGB stars. The median HB stellar mass was estimated at 0.61
M$_{\odot}$ for NGC 6752 by \cite{gratton10}. This was adopted by
\citetalias{lapenna16} and the current study. According to theory, lower
masses are possible. For example, \cite{dorman93} find a minimum envelope
mass for AGB ascension of 0.035~M$_{\odot}$, at the metallicity of NGC
6752. Adding their core mass of 0.48 M$_{\odot}$ suggests that the minimum
mass for an AGB star should be 0.52 M$_{\odot}$. This difference of $\sim
-0.1$~M$_{\odot}$ would systematically reduce the gravity of the AGB stars
by 0.07 dex. Importantly this would only affect AGB stars, leaving the RGB
gravities unchanged. However the AGB masses may also be higher. Assuming a
normal distribution of AGB masses around $0.61$~M$_{\odot}$ we thus
(roughly) estimate a $1\sigma$ error of $\sim \pm 0.05$ dex on the surface
gravity due to the uncertainty in total stellar mass. As discussed above
this can cause small changes in abundance results, particularly in the Fe
abundance derived from \ion{Fe}{ii}.

\subsubsection{Summary of uncertainties}
Here we have only explored some of the uncertainties inherent in
spectroscopic abundance determination. From this investigation it is clear
that, apart from the large uncertainty in $T_{\rm{eff}}$ given by some
colour-$T_{\rm{eff}}$ relations, and the uncertainties in measuring EWs,
there are many other sources of uncertainty that lead to additional
abundance differences of the order $0.01 \rightarrow 0.10$~dex. All the
uncertainties must combine, probably in a non-linear way, to create `noise'
and systematic shifts in the results, increasing the uncertainty of our
final abundances. It is also clear that each source of uncertainty
(including $T_{\rm{eff}}$) affects \ion{Fe}{i} and \ion{Fe}{ii} to
different degrees. It would thus be expected that each ion of each element
would also be affected to different degrees. For a broader view of
uncertainties in spectroscopic abundance determination we refer the
interested reader to the detailed empirical study of \cite{hinkel16}.

\section{Sodium from \citetalias{campbell13} data using new $T_{eff}$ scale}\label{sec:newsodium}

\begin{figure}
  \centering
  \includegraphics[width=\hsize]{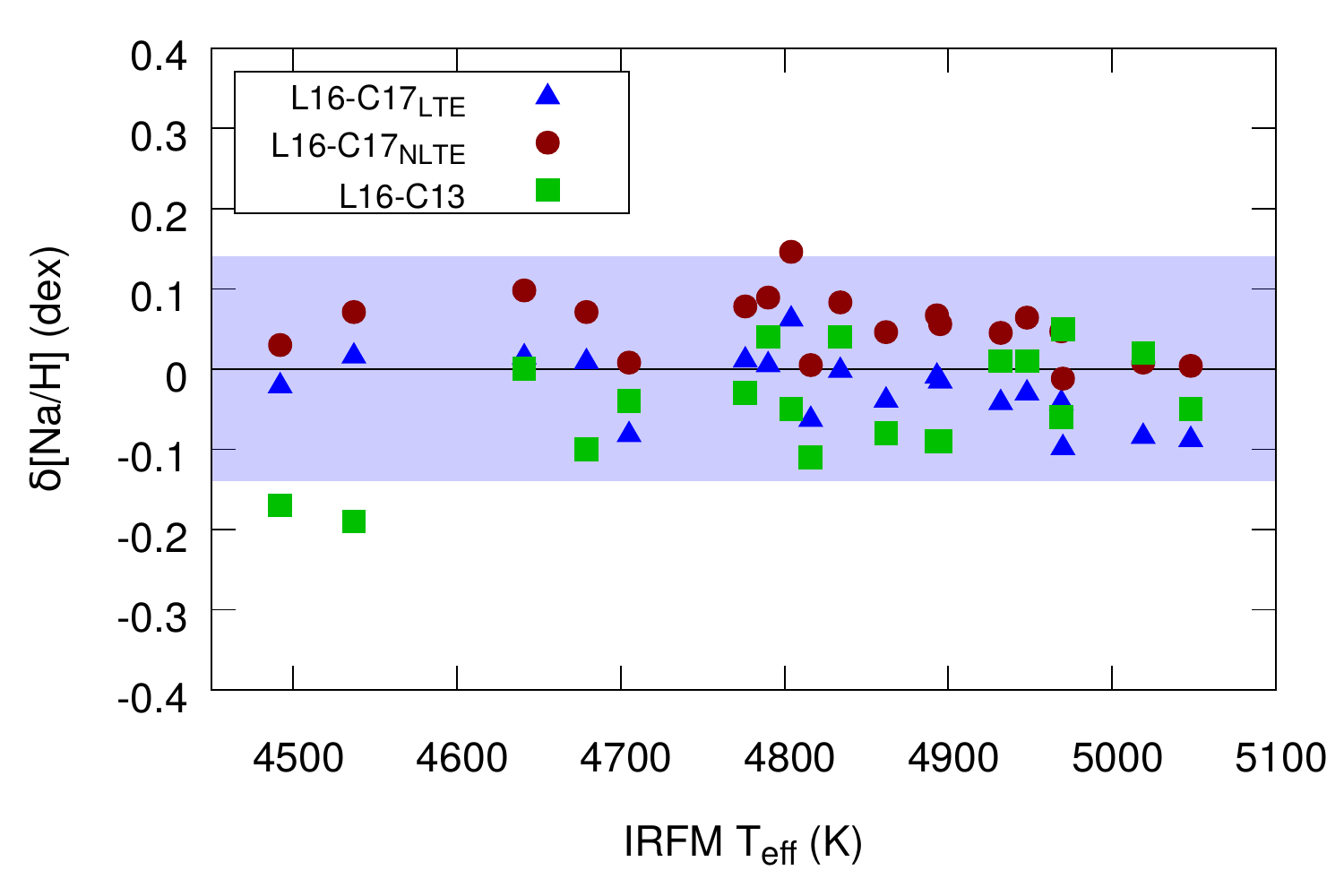}
  \caption{Comparison between our new AGB Na abundances (labelled C17), the
    \citetalias{campbell13} abundances, and those of
    \citetalias{lapenna16}. All abundance sets are corrected for NLTE
    effects except C17$_{\rm{LTE}}$, which is included to highlight the
    magnitude of the corrections. The shaded region denotes typical
    uncertainties of 0.1 dex, added in quadrature ($\sigma = 0.14$ dex in
    total). The two coolest stars are discussed in the text.}
  \label{fig:na-lte-compare}
\end{figure}

Of primary interest to the argument of \citetalias{campbell13} (and to a
lesser extent \citetalias{lapenna16}) is the distribution of sodium amongst
the AGB stars. It is possible that the new IRFM/$(V-K)$ temperature scale
(Fig.~\ref{fig:teffsAGB}) could remove the good agreement between [Na/H]
between studies (Fig.~\ref{fig:NaHcompare}), given that the offset is
$\sim 100$ K.

\ion{Na}{I} is predicted to suffer NLTE effects in giant stars such as
those studied here. \citetalias{campbell13} and \citetalias{lapenna16} both
used the \cite{gratton99} corrections to LTE abundances. For the current
study we have chosen to use the more recent NLTE corrections calculated by
\cite{lind11}. As noted by \cite{lind11} the \cite{gratton99} corrections
differ from most tabulations in the literature, especially at low
temperatures and gravities. The Na I lines we have used are: 5682.6~\AA,
5688.2~\AA, 6154.2~\AA, and 6160.7~\AA. The last two of these lines were
generally not detectable in the AGB stars, but a total of three or four
lines were usually detectable in the RGB stars.

We computed NLTE corrections for all available Na lines for all stars using
the INSPECT web interface\footnote{\url{http://www.inspect-stars.com}}
(\citealt{lind11}). The corrections were not large, with an average of
$-0.08$~dex in the AGB sample, and $-0.10$~dex for the RGB sample. In both
sets of stars this was essentially an offset, with the standard deviation
of the corrections being just $\sigma = 0.01$ dex (AGB) and 0.02 dex
(RGB). All corrections and final abundances are listed in
Table~\ref{tab:bigtab}.

In the RGB sample the average standard deviation of the line-to-line
abundance scatter, $\sigma_{ave}$, was reduced from 0.08 dex (LTE) to
0.06~dex~(NLTE). In the AGB stars $\sigma_{ave}$ reduced from 0.04~dex to
0.03~dex. Although these are small changes this is reassuring as it is what is
expected if the corrections are of the correct sign and magnitude.  
We note that since the AGB stars usually only have two lines measured, the
abundances in the AGB stars shouldn't be taken as more accurate than those
of the RGB stars. Indeed, we find the average line-to-line scatter
increases with the number of lines measured in the AGB stars, with
$\sigma_{ave} = 0.10$ dex in the stars with three detectable lines (the RGB
stars generally have three or four lines measurable). This is an important
point -- it is common practice to report standard deviations of very small
samples of lines as uncertainties in abundances. This can lead to
overconfidence in results.

In Figure \ref{fig:na-lte-compare} we compare our new [Na/H] values with
those of \citetalias{lapenna16} and \citetalias{campbell13}. Somewhat
surprisingly, it can be seen that practically all stars have essentially
the same abundances in all three studies, within the uncertainties. The
only exceptions are the two coolest stars: the current study and
\citetalias{lapenna16} find substantially higher Na than
\citetalias{campbell13} for these stars. This appears to be a minor error
in \citetalias{campbell13}, although it has no effect on the conclusions of
that study.

The average difference in Na abundance between \citetalias{lapenna16} and
the current study, in the sense \citetalias{lapenna16}$-$this study, is
$+0.05$~dex ($\sigma = 0.04$). Between \citetalias{lapenna16} and
\citetalias{campbell13} it is $-0.05$ dex ($\sigma = 0.07$), excluding the
2 coolest stars. It is interesting that the increase in temperature of
$\sim 100$~K (above the scale of \citetalias{lapenna16}) has no significant
effect on the Na abundances. This does concur with the observation that a
roughly 40~K temperature difference between \citetalias{campbell13} and
\citetalias{lapenna16} also had no significant effect
(Fig. \ref{fig:NaHcompare}). Adding to this that different tools, model
atmospheres, different stellar mass assumptions, and different spectra were
used between the studies strongly suggests that the Na abundances are very
robust, at least within the current analysis framework.

Our RGB sample has an average abundance of $\rm{[Na/H]}=-1.38$ dex, with a
standard deviation of $\sigma = 0.27$~dex. In contrast, the AGB sample has an
average abundance of $\rm{[Na/H]}=-1.71$ dex, with a standard deviation of
$\sigma = 0.13$~dex. While the spread in the RGB sample is much larger than
the uncertainties, the scatter in the AGB sample is low, and similar to the
0.10~dex reported by \citetalias{campbell13}. Indeed,
\citetalias{campbell13} noted that this level of scatter was similar to
their uncertainties and therefore consistent with a single abundance of
Na. We explore this topic further in the Discussion.

%--------------------------------------------------------------------
\section{Discussion and Conclusion}\label{sec:discuss}

We have investigated the differences between the NGC 6752 AGB star
abundance studies of \citetalias{campbell13} and \citetalias{lapenna16}. In
this paper we have focused on Fe and Na, since \citetalias{lapenna16}
reported a very strong apparent overionisation of \ion{Fe}{i}, and argued
that all neutral species, including \ion{Na}{i}, should be scaled by
\ion{Fe}{i}, thus altering the distribution of [Na/Fe] as compared to
\citetalias{campbell13} (Fig.~\ref{fig:NaFecompare}).

By dividing all neutral species by Fe derived from \ion{Fe}{I},
\citetalias{lapenna16} essentially made the assumptions that:
\begin{enumerate}
   \item Overionisation affects {\it all} neutral species
   \item They are affected by {\it exactly the same magnitude} of
overionisation
\end{enumerate}
While (1) is possible, (2) is highly unlikely, since the magnitudes of NLTE
effects are known to vary between species, and indeed between lines of the
same species. There are also variations with $T_{\rm{eff}}$, gravity, and
abundance of the element (see eg. \citealt{lind11,lind12} for the cases of
Na and Fe).

The scaling to \ion{Fe}{i} for neutral species in GC AGB stars was
originally suggested by \cite{ivans01}. They however proposed it as one of
{\it three} options to reconcile \ion{Fe}{i} and \ion{Fe}{ii} in the AGB
stars of the globular cluster \object{M5}. These options were arrived at
after an extensive investigation into the many possible causes of the Fe
discrepancy. The rationale behind the \ion{Fe}{i} scaling option was due to
the observation that: ``Whenever Fe appears to be overionized ... Si, Ti,
and V are excessively ionized by essentially the same amount...''
(\citealt{ivans01}). This was then generalised to include {\it all neutral
  species}. As \cite{ivans01} state, this was an extrapolation -- it was
not based on any theoretical calculations, since they weren't available at
that time. These calculations are now available, in particular for
\ion{Fe}{I} and \ion{Na}{i}. In the case of \ion{Fe}{i} the predicted NLTE
corrections (Sec.~\ref{sec:nlte}) are much smaller than the depression of
\ion{Fe}{i} reported by \citetalias{lapenna16}
(Fig.~\ref{fig:FeIdepression}). This has created a `tension' between
theory and observations (see also \citealt{lapenna14}).

During our comparison tests we immediately reproduced a strong apparent
depression of \ion{Fe}{i}, similar to
\citetalias{lapenna16}. Significantly, this was achieved with different
input data and computational tools. However further investigation showed
that this is primarily driven by the adopted stellar temperatures. By
deriving more reliable temperatures specifically for our stellar sample via
the IRFM method we found that the putative \ion{Fe}{I} overionisation
became insignificant. By comparing temperatures derived with colour-Teff
relations to our IRFM temperatures it was found that $(V-K)$ relations are
the most reliable (they have a much smaller uncertainties than $(B-V)$
relations), and we suggest that they be used for future (early) AGB
studies. Previous AGB studies that have reported strong apparent
\ion{Fe}{I} overionisation will need to be checked.

Interestingly, one of the possibilities canvassed by \cite{ivans01} to
remove the Fe discrepancy was indeed to ``arbitrarily increase the values
of $T_{\rm{eff}}$ above the Alonso et al. scale by 60~K on the RGB and
120~K on the AGB''. By investigating other options for temperature scales
they found that there were scales that were systematically hotter (see
their Section 3.3 for details). They concluded that this avenue to resolve
the Fe discrepancy ``remains an option''. 

Related to this is the study of 47 Tuc by \cite{lapenna14}. Noting that
their finding of a strong depression of \ion{Fe}{i} in the AGB stars was at
odds with theoretical predictions, they made a detailed investigation into
the uncertainties affecting $\delta\rm{Fe}$. They excluded $T_{\rm{eff}}$
as a significant source of uncertainty because the excitation balance was
``well satisfied'' in their sample. Our result -- that spectroscopically
determined temperatures tend to lie close to the initial estimates (usually
photometric) -- may be the reason that \cite{lapenna14} could not reconcile
the 47 Tuc AGB \ion{Fe}{i} and \ion{Fe}{ii} abundances. An indication that
this may be the case is given by one of the tests performed by
\cite{lapenna14}. They found that $\delta\rm{Fe}$ became negligible if
parameters from a higher mass star (1.2~M$_{\odot}$) were used. The
$T_{\rm{eff}}$ increase above the nominal value was $\sim +100$~K. The fact
that this was a positive $T_{\rm{eff}}$ offset that reduced $\delta\rm{Fe}$
matches with our findings (Fig.~\ref{fig:tefftest}). The magnitude of this
$T_{\rm{eff}}$ offset is similar to the offsets we have found here
(Fig.~\ref{fig:teffsAGB}), so we would expect a greater increase in
\ion{Fe}{i}, of the order $\sim 0.2$~dex as compared to $\sim0.1$~dex found
by \cite{lapenna14}. However the much higher metallicity of 47 Tuc may
reduce the effect.

\citetalias{lapenna16} also discussed their results in terms of [X/H],
noting that using Fe in the denominator may have skewed the results (they
report that it did not). However the different $T_{\rm{eff}}$ scale
presented here\footnote{Gravity should be only slightly affected, with
  changes most likely smaller than the \citetalias{lapenna16} uncertainty
  of $\sim 0.1$~dex (Sec.~\ref{sec:FeC13}).} means that all elements need
reanalysis, since the effects are uncertain and most likely vary from
element to element, as our investigation of Fe and Na has shown. As
molecular band formation is highly dependent on temperature, abundances for
elements based on molecules (C, N, and indirectly, O, in
\citetalias{lapenna16}) are likely to be altered. We are unsure how the
results of the \citetalias{lapenna16} study will be affected.  The
  effects may or may not alter the conclusions on the topic of AGB
  subpopulations, but it clearly requires investigation.

After the reconciliation of \ion{Fe}{i} and \ion{Fe}{ii}, we checked the
effect of the new temperature scale on derived sodium abundances and found
that it had no significant effect. Interestingly the [Na/H] abundances
across the three studies (\citetalias{campbell13}, \citetalias{lapenna16},
and the current study) all agree, within the uncertainties. It is also
remarkable that such large differences in temperatures, gravity, input
data, and tools, have such little effect on the Na abundances. This
suggests they are very robust -- at least in the current paradigm of 1D
stellar atmosphere abundance determinations and NLTE modelling. We
speculate that either (i) significant changes in the (currently small)
predicted NLTE corrections, or (ii) 3D atmosphere effects, would be the most
likely factors that could alter the Na abundances. Since Fe is constant in
\object{NGC 6752} then [Na/Fe] must also be consistent between all
studies, aside from possible systematic offsets in Fe.

We found that the Na abundance spread in the AGB sample is slightly larger
than that found by \citetalias{campbell13} ($\sigma = 0.13$ dex versus 0.10
dex). This compares with a spread of $\sigma = 0.27$ dex in the RGB
sample. As shown in \citetalias{campbell13} the AGB distribution is
centered over the RGB SP1 distribution. However we are cautious to assign
significance to the slight increase in the AGB dispersion, especially given
the exploration of the many sources of uncertainty in
Section~\ref{sec:uncerts}. In particular we note the uncertainties in NLTE
corrections that could amount to 0.1~dex, but which are generally ignored
in abundance studies.  We also recall that the Na abundances in the AGB
stars are usually based on just two lines, so the line-to-line scatter
uncertainty is not well constrained. Indeed, we found that the line-to-line
scatter {\it increases} in the stars with more Na lines measured, to a an
average value of 0.10 dex in the AGB stars with three detectable
lines\footnote{This compares with the LTE (NLTE) average scatter of 0.08
  (0.06) dex in the RGB stars, which mostly have three or four lines
  measured.}. As this is just one source of uncertainty, it should be taken
as a {\it minimum} for the total abundance uncertainty. That said, our
empirical result in Figure~\ref{fig:na-lte-compare}, which shows that the
Na abundances appear very robust between the three studies (all within $\pm
0.05$ dex, on average), may suggest that the spread is
significant\footnote{Although it may instead reflect that usually the same
  two Na lines are used in every AGB study, and, as we have shown, that
  these particular lines are insensitive to many factors.}. If it is, this
could be considered a small signal of the tail of the second population
distribution being present on the AGB, similar to that found for
\object{M4} (see Discussion in \citealt{maclean16}). However, given the
small magnitude of the signal, it is advisable to attempt to identify the
subpopulations in multi-dimensional chemical space instead, as noted by
others (eg. \citetalias{lapenna16}; \citealt{maclean16}). Thus we leave the
further discussion of AGB subpopulations to our next paper in the series,
where we will investigate other elements given the improved AGB
temperatures. What appears certain is that the Na spread in the AGB stars
is very restricted compared to the RGB stars.

We conclude by noting that care must be taken in deriving AGB star
temperatures, and, more generally, that uncertainty reporting in abundance
analysis papers should be more robust -- there are many sources of
uncertainty that can significantly alter the results. The relevant example
here is the standard procedure of testing the abundance results'
sensitivity to temperature. Both \citetalias{campbell13} and
\citetalias{lapenna16} used an estimated uncertainty of $\sim \Delta
T_{\rm{eff}} \pm 30 K$. This uncertainty is small compared to the
$T_{\rm{eff}}$ differences of $\sim 60-100$ K found here. If more realistic
uncertainties were included, the error bars on \ion{Fe}{i} would have been
large\footnote{See final paragraph of Section~\ref{sec:A97}.} -- and the
results consistent with theory and other abundance studies. 

%%==================================

\begin{acknowledgements}
The spectroscopic observations were carried out with the Very Large
Telescope under ESO programme 089.D-0038 (PI Campbell). Part of this work
was supported by the DAAD (PPP project 57219117) with funds from the German
Federal Ministry of Education and Research.  DY acknowledges funding from
the Australian Research Council (FT140100554). BTM acknowledges the
financial support of the Australian Postgraduate Award scholarship. We
thank Yazan Momany and Frank Grundahl for providing their photometric
datasets, and Yue Wang and Francesca Primas for helpful discussions. This
work made extensive use of the SIMBAD, Vizier, 2MASS, and NASA ADS
databases. It also made use of the open-source software packages
MOOG (Chris Sneden; \url{http://www.as.utexas.edu/~chris/moog.htmland}) and
$q2$ (Ivan Ramirez; \url{https://github.com/astroChasqui/q2}). We thank the
anonymous referee for their constructive comments, which improved the paper.
\end{acknowledgements}

%%==================================

\bibliographystyle{aa}
\bibliography{decra}

%%==================================

%\let\cleardoublepage\clearpage

\renewcommand{\cleardoublepage}{}
\appendix %Table 2 moved to appendix as requested by A&A.
\section{}
\begin{table*}
%
%\begin{table}[]
\centering
\caption{Final stellar parameters and abundance results.}
\label{tab:bigtab}
\begin{tabular}{lc|ccc|cc|cc|cc|c|cc}
\hline\hline
ID     & Type & $T_{\rm{eff}}$ & $\log g$ & Xi   & \ion{Fe}{I} & $\sigma$ &\ion{Fe}{II} & $\sigma$ & $\ion{Na}{I}_{\rm{LTE}}$ & $\sigma$ & Corr$_{\ion{Na}{I}}$ & $\ion{Na}{I}_{\rm{NLTE}}$ &
$\sigma$\\
      &       &  (K)         &   (dex)  & (km/s) & (dex)     &  (dex)
&(dex)   & (dex)   & (dex) & (dex) & (dex) & (dex) & (dex)\\ 
\hline
22     & AGB  & 4641 & 1.31 & 1.80 & 6.01 & 0.11       & 5.92 & 0.06        & 4.83     & 0.06            & -0.08           & 4.75     & 0.05             \\
25     & AGB  & 4492 & 1.09 & 1.87 & 6.07 & 0.10       & 6.03 & 0.11        & 4.56     & 0.10            & -0.05           & 4.51     & 0.11             \\
31     & AGB  & 4537 & 1.20 & 1.83 & 6.08 & 0.11       & 5.99 & 0.09        & 4.52     & 0.04            & -0.05           & 4.47     & 0.03             \\
44     & AGB  & 4679 & 1.43 & 1.76 & 6.03 & 0.12       & 6.03 & 0.13        & 4.52     & 0.01            & -0.06           & 4.46     & 0.00             \\
52     & AGB  & 4862 & 1.68 & 1.68 & 5.96 & 0.10       & 6.01 & 0.16        & 4.70     & 0.04            & -0.08           & 4.61     & 0.03             \\
53     & AGB  & 4790 & 1.57 & 1.72 & 5.99 & 0.08       & 6.01 & 0.07        & 4.76     & 0.03            & -0.08           & 4.68     & 0.02             \\
59     & AGB  & 4804 & 1.61 & 1.70 & 5.92 & 0.10       & 5.94 & 0.08        & 4.71     & 0.03            & -0.08           & 4.62     & 0.02             \\
60     & AGB  & 4776 & 1.57 & 1.72 & 6.12 & 0.12       & 6.05 & 0.12        & 4.54     & 0.00            & -0.07           & 4.47     & 0.01             \\
61     & AGB  & 4834 & 1.63 & 1.70 & 6.04 & 0.10       & 6.01 & 0.09        & 4.73     & 0.04            & -0.08           & 4.65     & 0.02             \\
65     & AGB  & 4705 & 1.43 & 1.76 & 6.00 & 0.10       & 6.02 & 0.05        & 4.95     & 0.05            & -0.09           & 4.86     & 0.05             \\
75     & AGB  & 4816 & 1.67 & 1.68 & 5.91 & 0.07       & 6.05 & 0.11        & 4.50     & 0.01            & -0.07           & 4.44     & 0.00             \\
76     & AGB  & 4970 & 1.76 & 1.65 & 5.92 & 0.11       & 5.97 & 0.09        & 4.89     & 0.14            & -0.08           & 4.80     & 0.15             \\
78     & AGB  & 4948 & 1.76 & 1.65 & 5.93 & 0.12       & 5.97 & 0.10        & 4.86     & 0.04            & -0.09           & 4.77     & 0.02             \\
80     & AGB  & 4893 & 1.75 & 1.66 & 5.95 & 0.09       & 6.07 & 0.07        & 4.59     & 0.04            & -0.07           & 4.51     & 0.03             \\
83     & AGB  & 4932 & 1.76 & 1.65 & 5.96 & 0.09       & 6.03 & 0.15        & 4.71     & 0.01            & -0.09           & 4.63     & 0.00             \\
94     & AGB  & 4969 & 1.83 & 1.63 & 5.90 & 0.07       & 6.08 & 0.10        & 4.73     & 0.07            & -0.09           & 4.64     & 0.06             \\
97     & AGB  & 5048 & 1.89 & 1.61 & 5.92 & 0.10       & 6.06 & 0.08        & 4.78     & 0.02            & -0.09           & 4.69     & 0.03             \\
104    & AGB  & 4895 & 1.79 & 1.64 & 5.94 & 0.10       & 6.07 & 0.19        & 4.48     & 0.00            & -0.07           & 4.41     & 0.00             \\
201620 & AGB  & 5019 & 1.83 & 1.63 & 6.01 & 0.11       & 6.01 & 0.04        & 4.84     & 0.01            & -0.09           & 4.75     & 0.02             \\
12     & RGB  & 4348 & 1.10 & 1.87 & 6.09 & 0.12       & 5.92 & 0.09        & 5.10     & 0.05            & -0.08           & 5.01     & 0.07             \\
23     & RGB  & 4404 & 1.19 & 1.84 & 6.09 & 0.10       & 6.12 & 0.02        & 5.12     & 0.08            & -0.09           & 5.04     & 0.04             \\
27     & RGB  & 4453 & 1.30 & 1.80 & 6.08 & 0.13       & 6.13 & 0.04        & 4.77     & 0.18            & -0.06           & 4.70     & 0.18             \\
29     & RGB  & 4362 & 1.13 & 1.86 & 6.07 & 0.11       & 6.07 & 0.17        & 4.67     & 0.04            & -0.06           & 4.61     & 0.05             \\
30     & RGB  & 4362 & 1.10 & 1.87 & 6.09 & 0.12       & 5.93 & 0.11        & 5.10     & 0.18            & -0.09           & 5.01     & 0.13             \\
35     & RGB  & 4490 & 1.37 & 1.78 & 6.09 & 0.12       & 6.05 & 0.05        & 5.39     & 0.13            & -0.11           & 5.28     & 0.08             \\
43     & RGB  & 4469 & 1.37 & 1.78 & 6.04 & 0.09       & 5.99 & 0.07        & 5.45     & 0.11            & -0.12           & 5.33     & 0.04             \\
50     & RGB  & 4436 & 1.27 & 1.81 & 6.11 & 0.12       & 6.08 & 0.16        & 4.95     & 0.06            & -0.09           & 4.86     & 0.03             \\
54     & RGB  & 4571 & 1.51 & 1.73 & 6.18 & 0.11       & 6.08 & 0.15        & 4.94     & 0.03            & -0.09           & 4.85     & 0.03             \\
64     & RGB  & 4467 & 1.36 & 1.78 & 6.07 & 0.09       & 5.98 & 0.11        & 5.39     & 0.15            & -0.12           & 5.27     & 0.08             \\
91     & RGB  & 4641 & 1.75 & 1.66 & 5.99 & 0.12       & 6.01 & 0.12        & 5.10     & 0.06            & -0.09           & 5.01     & 0.06             \\
92     & RGB  & 4672 & 1.73 & 1.66 & 6.12 & 0.11       & 6.02 & 0.13        & 4.72     & 0.11            & -0.08           & 4.64     & 0.12             \\
107    & RGB  & 4681 & 1.83 & 1.63 & 6.07 & 0.12       & 5.99 & 0.10        & 5.03     & 0.08            & -0.09           & 4.94     & 0.09             \\
129    & RGB  & 4738 & 1.94 & 1.60 & 6.03 & 0.12       & 6.08 & 0.20        & 5.04     & 0.05            & -0.10           & 4.94     & 0.06             \\
155    & RGB  & 4752 & 2.00 & 1.58 & 6.02 & 0.13       & 6.01 & 0.09        & 4.62     & 0.01            & -0.08           & 4.54     & 0.02             \\
161    & RGB  & 4852 & 2.07 & 1.55 & 6.14 & 0.09       & 6.05 & 0.10        & 5.23     & 0.06            & -0.12           & 5.11     & 0.04             \\
170    & RGB  & 4792 & 2.07 & 1.55 & 6.07 & 0.14       & 6.05 & 0.08        & 5.40     & 0.08            & -0.12           & 5.27     & 0.02             \\
186    & RGB  & 4832 & 2.13 & 1.54 & 6.02 & 0.15       & 5.99 & 0.13        & 4.63     & 0.06            & -0.08           & 4.54     & 0.04             \\
193    & RGB  & 4854 & 2.14 & 1.53 & 6.06 & 0.14       & 6.11 & 0.08        & 4.58     & 0.00            & -0.08           & 4.50     & 0.01             \\
262    & RGB  & 4873 & 2.25 & 1.50 & 6.03 & 0.09       & 5.92 & 0.04        & 5.17     & 0.04            & -0.12           & 5.06     & 0.01             \\
276    & RGB  & 4864 & 2.25 & 1.50 & 5.99 & 0.09       & 6.02 & 0.08        & 5.20     & 0.10            & -0.12           & 5.08     & 0.06             \\
200619 & RGB  & 4731 & 1.91 & 1.61 & 6.02 & 0.07       & 6.06 & 0.09        & 5.41     & 0.01            & -0.08           & 5.33     & 0.01            \\
\hline
\end{tabular}
\tablefoot{Abundances are presented as $\log\epsilon =
  \log(\rm{N_{X}/N_{H}}) + 12$, where X represents each species. The
  $\sigma$ values are based on line-to-line abundance scatter only. The NLTE
  corrections to the LTE sodium abundances are given in the third last column. }
\end{table*}

\end{document}